\documentclass[acmsmall,screen,nonacm]{acmart}

\usepackage[T1]{fontenc}
\usepackage[utf8]{inputenc}
\usepackage{fontawesome5}

\usepackage{mathpartir}
\usepackage{graphicx}
\usepackage{makecell}
\usepackage{hyperref}
\usepackage{multicol}
\usepackage{stmaryrd} %
\usepackage{xspace}
\usepackage{xcolor}
\usepackage{amsfonts}
\usepackage{adjustbox}
\usepackage{soul}
\usepackage{graphicx}
\usepackage{multirow}

\usepackage{minted}
\newlength{\mgcwidth}

\newmintinline[ocamli]{ocaml}{}

\newif\ifshort
\ifdefined\shortversion\shorttrue\fi
\ifshort
\usepackage{xr}
\fi

\usepackage{subcaption}
\usepackage[nameinlink,noabbrev]{cleveref}
\crefname{section}{\S\!}{\S}
\crefformat{section}{\S#2#1#3}
\Crefformat{section}{Section~#2#1#3}
\Crefname{section}{Section}{Sections}
\crefname{figure}{Figure}{Figures}
\crefname{theorem}{Theorem}{Theorems}
\crefname{corollary}{Corollary}{Corollaries}

\ifshort
\newcommand{\appref}[1]{\citep[\cref*{long-#1}]{anonymous-supplementary-material}}
\newcommand{\Appref}[1]{\Cref*{long-#1}~\citep{anonymous-supplementary-material}}
\else
\newcommand{\appref}[1]{\cref{#1}}
\newcommand{\Appref}[1]{\Cref{#1}}
\fi

\newtheorem{theorem}{Theorem}[section]

\usepackage{thmtools}
\usepackage{thm-restate}

\usepackage{iris}

\usepackage{xspace}
\usepackage{expl3}

\newcommand{\captionlabel}[2]{%
  \vspace{-1em}%
  \caption{#1}%
  \Description{#1}%
  \label{#2}%
  \vspace{-1em}%
}

\newcommand{\zoolang}{ZooLang\xspace}
\newcommand{\lang}{MizzleLang\xspace}
\newcommand{\logic}{Mizzle\xspace}

\newcommand{\ocaml}{OCaml~5\xspace}

\renewcommand{\star}{\ast}
\newcommand{\wider}[1]{\,#1\,}
\newcommand{\morespacingaroundstar}{%
\let\oldstar\star
\renewcommand{\star}{\wider\oldstar}%
}
\newcommand{\morespacingaroundwedge}{%
\let\oldwedge\wedge
\renewcommand{\wedge}{\wider\oldwedge}%
}
\newcommand{\morespacingaroundvdash}{%
\let\oldvdash\vdash
\renewcommand{\vdash}{\wider\oldvdash}%
}
\newcommand{\mlprec}[1]{\left(\,\begin{array}[c]{@{}r@{}}#1\end{array}\,\right)}
\newcommand{\prowstrut}{\vphantom{\begin{array}[c]{@{}c@{}}X\\X\end{array}}}
\newcommand{\pure}[1]{\ulcorner #1 \urcorner}

\newcommand{\iTrue}{\top}
\newcommand{\pre}{\varphi}
\newcommand{\post}{\psi}

\newcommand{\bigast}[2]{{\scaleobj{2}{\ast}}_{#1}\,#2}

\newcommand{\ofs}{i}

\newcommand{\funcname}{f}
\newcommand{\body}{\expr}
\newcommand{\argname}{x}

\newcommand{\disjoint}[2]{#1\,\cap\,#2 = \emptyset}

\newcommand{\config}{c}
\newcommand{\mkconf}[4]{#1 \,/\, #2 \,/\, #3 \,/\, #4}

\newcommand{\trace}{\theta}
\newcommand{\step}[2]{#1 \rightarrow #2}

\newcommand{\steprtc}[2]{#1 \rightarrow^{\star} #2}

\newcommand{\confinit}[1]{\textsf{init}(#1)}
\newcommand{\locs}[1]{\textsf{locs}(#1)}

\newcommand{\threadname}{\textsf{thread}\xspace}
\newcommand{\thread}[3]{\threadname\,#1\,#2\,#3}
\newcommand{\remainingname}{\textsf{remaining}\xspace}
\newcommand{\remaining}[1]{\remainingname\,#1}
\newcommand{\goal}{\mathfrak{goal}}
\newcommand{\tid}{\pi}

\newcommand{\nilctx}{\epsilon}

\newcommand{\cells}[2]{#1 \mathrel{\mapsto} #2}

\newcommand{\headername}{\textsf{header}\xspace}
\newcommand{\header}[3]{\headername\,#1\,#2\,#3}
\newcommand{\sizeof}[1]{\lvert #1 \rvert}

\newcommand{\istrace}[1]{\textsf{trace}\,#1}
\newcommand{\vread}[2]{#1(#2)}

\newcommand{\stuckname}{\textsf{Stuck}\xspace}
\newcommand{\stuck}[1]{\stuckname\,#1}

\newcommand{\ectx}{\kappa}
\newcommand{\ectxs}{\vec\ectx}

\newcommand{\bigsep}{\mathop{\vcenter{\hbox{\large$\star$}}}}
\newcommand{\parfor}{\textsf{parallel\_for}}

\newcommand{\eif}[3]{\textsf{if}\,#1\,\textsf{then}\,#2\,\textsf{else}\,#3}

\newcommand{\eassert}[1]{\textsf{assert}\,#1}
\newcommand{\vtrue}{\textsf{true}}
\newcommand{\vfalse}{\textsf{false}}
\newcommand{\vunit}{()}

\newcommand{\wal}{w}
\newcommand{\vbool}{b}
\newcommand{\vint}{i}
\newcommand{\loc}{\ell}
\newcommand{\tagn}{n}

\newcommand{\Values}{\textit{Val}}
\newcommand{\Bool}{\mathbb{B}}
\newcommand{\Int}{\mathbb{Z}}

\newcommand{\atm}{\alpha}
\newcommand{\matomic}{\textsf{at}}
\newcommand{\mnonatomic}{\textsf{na}}

\newcommand{\unop}{\ominus}
\newcommand{\binop}{\oplus}

\newcommand{\vrec}[3]{\overline{\textsf{rec}}\,#1\,#2 = #3}

\newcommand{\erec}[3]{\textsf{rec}\,#1\,#2 = #3}
\newcommand{\eapp}[2]{#1\,#2}
\newcommand{\elet}[3]{\textsf{let}\,#1 = #2\,\textsf{in}\,#3}
\newcommand{\eunop}[2]{#1\,#2}
\newcommand{\ebinop}[3]{#1 \mathbin{#2} #3}
\newcommand{\eequal}[2]{#1 \mathrel{\mathtt{==}} #2}

\newcommand{\eblock}[2]{\textsf{block}\,#1\,#2}

\newcommand{\egettag}[1]{\textsf{tag}\,#1}
\newcommand{\egetsize}[1]{\textsf{size}\,#1}
\newcommand{\ealloc}[2]{\textsf{alloc}\,#1\,#2}
\newcommand{\eload}[3][\atm]{#2.[#3]^{#1}}
\newcommand{\estore}[4][\atm]{#2.[#3]^{#1} \mathbin{\leftarrow} #4}

\newcommand{\ecas}[3]{\textsf{CAS}\,#1\,#2\,#3}

\newcommand{\vtuple}[1]{(#1)}
\newcommand{\locaccname}{\textsf{LAcc}\xspace}
\newcommand{\locacc}[4]{\locaccname\,#1\,#2\,#3\,#4}
\newcommand{\efork}[1]{\textsf{fork}\,#1}
\newcommand{\eemit}[1]{\textsf{emit}\,#1}

\newcommand{\khole}{\square}
\newcommand{\concat}{\mathbin{+\!\!+}}

\newcommand{\Oob}{Out-of-bound\xspace}
\newcommand{\oob}{out-of-bound\xspace}

\newcommand{\store}{\sigma}
\newcommand{\tagmap}{\tau}
\newcommand{\exprs}{\vec{\expr}}
\newcommand{\mapping}{\psi}
\newcommand{\headerval}[2]{(#1, #2)}
\newcommand{\mapupd}[3]{[#1 \mathbin{:=} #2]#3}

\newcommand{\headto}{\mathrel{\rightarrow_{\textsf{h}}}}
\newcommand{\threadto}{\mathrel{\rightarrow_{\textsf{t}}}}
\newcommand{\headstep}[9]{\mkconf{#1}{#3}{#2}{#4} \headto \mkconf{#5}{#7}{#6}{#8}\; ;\; #9}
\newcommand{\threadstep}[9]{\mkconf{#1}{#3}{#2}{#4} \threadto \mkconf{#5}{#7}{#6}{#8}\; ;\; #9}
\newcommand{\fillk}[2]{#1[#2]}

\newcommand{\interpname}{\textsf{interp}\xspace}
\newcommand{\interp}[1]{\interpname\,#1}

\newcommand{\racyname}{\textsf{Racy}\xspace}
\newcommand{\racy}[1]{\racyname\,#1}

\newcommand{\witnessname}{\textsf{witness}\xspace}
\newcommand{\witness}[1]{\witnessname\,#1}

\newcommand{\isracy}{\witness\racyname}

\newcommand{\caccname}{\textsf{Acc}\xspace}
\newcommand{\cacc}[4]{\caccname\,#1\,#2\,#3\,#4}

\newcommand{\mirrorname}{\textsf{mirror}\xspace}
\newcommand{\mirror}[1]{\mirrorname\,#1}

\newcommand{\tr}[2]{\lfloor #2 \rfloor_{#1}}
\newcommand{\mirrorinvsname}{\textsf{mirror\_invs}\xspace}
\newcommand{\mirrorinvs}[4]{\mirrorinvsname\,#1\,#2\,#3\,#4}
\newcommand{\loctokname}{\textsf{live}\xspace}
\newcommand{\loctok}[1]{\loctokname\,#1}

\newcommand{\defeq}{\triangleq}
\newcommand{\optag}{\tau}
\newcommand{\evcall}[1]{\textsf{Call}\,#1}
\newcommand{\evlin}[2]{\textsf{Lin}\,#1\,#2}
\newcommand{\evret}[1]{\textsf{Ret}\,#1}
\newcommand{\tagged}[2]{(#1, #2)}
\renewcommand{\trace}{\theta}
\newcommand{\restrict}[2]{#1 {\restriction}_{#2}}
\newcommand{\proj}[2]{\restrict{#1}{#2}}
\newcommand{\delin}[1]{\restrict{#1}{\textsf{call},\textsf{ret}}}
\newcommand{\onlin}[1]{\restrict{#1}{\textsf{lin}}}
\newcommand{\prefixeq}{\sqsubseteq}
\newcommand{\Model}{\mathcal{M}}
\newcommand{\Sset}{\mathbb{S}}
\newcommand{\sinit}{s_0}
\newcommand{\valreq}[1]{\textsf{Valid}\,#1}
\newcommand{\reqresp}[4]{\smash{#1 \mathrel{\xrightarrow{\smash{\raisebox{-1pt}{$\scriptstyle #2\,\Rightarrow\,#3$}}}} #4}}
\newcommand{\soundlinname}{\textsf{SoundLin}\xspace}
\newcommand{\drives}[3]{\soundlinname\,#2\,#1\,#3}
\newcommand{\callvalid}[1]{\textsf{ValidCalls}\,#1}
\newcommand{\pertaglin}[1]{\textsf{PerTag}\,#1}
\newcommand{\linof}[2]{\textsf{Linearizes}\,#1\,#2}
\newcommand{\nonlinname}{\textsf{NonLinearizable}\xspace}
\newcommand{\nonlin}[1]{\nonlinname\,#1}

\renewcommand{\wp}[2]{\textsf{wp}\,#1\,#2}

\newcounter{remark}[section]

\ExplSyntaxOn
\newcommand\latinabbrev[1]{
  \peek_meaning:NTF . {%
    \emph{#1}\@}%
  { \peek_catcode:NTF a {%
      \emph{#1}.\@\xspace}%
    {\emph{#1}.\@\xspace}}}
\ExplSyntaxOff

\def\eg{\latinabbrev{e.g}}

\def\etc{\latinabbrev{etc}}
\def\ie{\latinabbrev{i.e}}

\renewcommand{\TirNameStyle}[1]{\hypertarget{#1}{\textsc{#1}}}
\newcommand{\RULE}[1]{\hyperlink{#1}{\textsc{#1}}\xspace}

\makeatletter
\def\arcr{\@arraycr}
\makeatother

\setcitestyle{nosort}

\begin{document}

\title{Mizzle: A Complete Concurrent Incorrectness Logic with Applications to Agentic Bug Certification}

\author{Alexandre Moine}
\orcid{0000-0002-2169-1977}
\email{alexandre.moine@inria.fr}
\affiliation{%
  \institution{Inria, CNRS, ENS de Lyon, Université Lyon 1, LIP}
  \city{Lyon}
  \country{France}
}
\authornote{Part of this work was done while affiliated with New York University.}

\author{Sam Westrick}
\orcid{0000-0003-2848-9808}
\email{shw8119@nyu.edu}
\affiliation{%
  \institution{New York University}
  \city{New York}
  \country{USA}
}

\author{Joseph Tassarotti}
\orcid{0000-0001-5692-3347}
\email{jt4767@nyu.edu}
\affiliation{%
  \institution{New York University}
  \city{New York}
  \country{USA}
}
\authornote{Also affiliated with Amazon Web Services. This paper does not reflect the views of Amazon Web Services.}

\begin{abstract}
Large language models are increasingly used
to find bugs in real-world programs,
but they also produce a flood of false alarms
that waste developers' time.
We propose to prevent these false alarms
by requiring an LLM to accompany each bug report
with a machine-checked proof, in a program logic,
that the reported bug is real.
Unlike a test, such a proof is unaffected
by how rarely a concurrency bug manifests,
and can certify a bug against a library's specification
rather than its implementation.
We follow the approach of incorrectness logics,
whose under-approximate reasoning establishes
that a claimed behavior is genuinely reachable,
and hence a true positive.
In our case, however,
the logic must model a realistic programming language,
be mechanized so that proofs can be checked,
and be complete, so that no real bug is ruled out
for want of a derivation.

We present Mizzle, an incorrectness separation logic
for concurrent programs written in a substantial subset of OCaml,
parametric in the notion of incorrectness.
We mechanize Mizzle in the Rocq proof assistant
on top of the Iris framework,
and we prove that it is both sound,
so that it never justifies a false alarm,
and complete, so that every incorrect execution admits a derivation.
We instantiate Mizzle with three notions of incorrectness:
stuckness (triggering undefined behavior),
the presence of a memory race (in a sequentially consistent setting),
and the non-linearizability of a data structure.
We show that an LLM agent can use Mizzle
to certify seeded memory races and linearizability bugs
for under a dollar per bug on average.
We also report two previously unknown linearizability bugs
in an OCaml library of concurrent data structures.

\end{abstract}

\maketitle

\section{Introduction}
\label{sec:intro}
False alarms in bug-finding tools waste developers' time.
As a result, some bug-finding tools are designed to avoid all false alarms and only report \emph{true positives}: when they claim there is a bug, then the program really exhibits the buggy behavior.
But how can one prove that the analyses conducted by these tools really result in only true positives?
Traditionally, doing so required ad-hoc proofs that could be quite challenging.
An alternative that has emerged in recent years is to use \emph{Incorrectness Logic}~\citep{ohearn-il-20} as a formal foundation for justifying that a program analysis only yields true positives.
Whereas traditional Hoare logic over-approximates a program's behaviors, so that specifications show that a program's behaviors must belong to some set, incorrectness logics instead use \emph{under}-approximation, so that a specification shows that some set of behaviors is a subset of a program's behaviors; \ie, they are actual observable behaviors and not false alarms.
Incorrectness logics have been successfully applied to justify a number of static analyses and bug-finding methods~\citep{raad-isl-20, le-et-al-22, raad-et-al-22, raad-et-al-23}.

Today, developers are awash in bug reports coming from a whole new class of bug-finding tools: large language models (LLMs).
Unfortunately, just as with traditional bug-finding tools, false alarms from LLMs \emph{also} waste developers' time.
Indeed, developers have become so inundated with spurious reports from LLMs that some popular open source projects like \texttt{curl} have announced closing their bug bounty programs~\citep{curl-bug-bounty-end}.
At the same time, LLMs are remarkably good at finding subtle and serious bugs that are beyond the scope of traditional methods.
Thus, a natural question is whether we can find a way to retain the benefits of LLM bug finding while reducing the incidence of false alarms.
Today, one popular method for doing so involves throwing yet more LLMs at the problem, often by asking multiple models to scrutinize a bug report, which offers no guarantees.
Another approach is to require the LLM to construct a reproducible test case that triggers the bug.
But concurrency bugs might only be triggered under rare schedules, making them hard to tell from false alarms~(\cref{sec:study:saturn}).
Controlled schedulers can replay such a schedule deterministically,
but a replay only certifies the bug against the current implementation of every component involved,
missing behaviors that a library specification permits~(\cref{sec:key:logic}).

This paper proposes an alternative approach of using incorrectness logic to certify that LLM bug reports are not false alarms.
LLMs are a black box, so unlike previous efforts to apply incorrectness logic to bug-finding tools, we cannot use the logic to prove that the LLM never returns a false alarm.
Instead, we shift the burden of proof to the LLM: with each bug report, it must produce a derivation in an incorrectness logic showing, step by step, how the bug arises.
In particular, by mechanizing the incorrectness logic inside of a proof assistant such as Rocq, and requiring the LLM to carry out a proof using that embedding, we can use the proof assistant to check that the proof is valid.

For this approach to be viable, we need an incorrectness logic with several properties.
First, it must handle programs written in a real programming language, modeling a substantial portion of that language's features.
Second, it must be mechanized, so that derivations can be checked.
Third, the logic ought to be \emph{complete}: if a bug exists, then it must be possible to construct a derivation in the logic showing so.
Otherwise, a bug might be impossible to exhibit in the logic.

To the best of our knowledge, no existing incorrectness logic for a language with concurrency has these properties,
because existing incorrectness logics for concurrency address a different problem.
For example, CISL~\citep{raad-et-al-22},
designed to underpin automated bug finders,
and its generalization CASL~\citep{raad-et-al-23} are theories of \emph{thread-modular} under-approximate reasoning,
for analyses that detect bugs by reasoning about each thread in isolation.
But sometimes exhibiting a bug is most easily done in a non-modular way, by considering a particular concrete interleaving of several threads.
In CISL, following a particular interleaving is a non-compositional fallback,
supported only at the granularity of the program's sequential compositions;
CASL instead lets each thread assume actions of its environment,
at the price of rely-guarantee reasoning and recorded traces.
Moreover, these logics are not known to be complete,
and they target a simple imperative language:
it has neither functions, nor higher-order code, nor unstructured thread creation.
Thus, while these logics are a good fit for justifying the soundness of various forms of modular static analysis, they are not well-suited to the problem of certifying a bug proposed by an LLM.

For certifying a proposed bug,
the LLM supplies an interleaving,
and the certifier must confirm the existence of the bug with a checkable proof.
Modular reasoning can still be useful, for reasoning about parts of the program where the details of the interleaving are not relevant,
but we also need the ability to reason through the steps of that particular interleaving at points.
To address this, we present \logic,
a concurrent incorrectness separation logic for a substantial subset of \ocaml{}.
More precisely, we make the following contributions:
\begin{itemize}
\item \logic, an incorrectness separation logic for concurrent programs
written in a
subset of \ocaml under a sequential consistency model,
parametric in the notion of incorrectness.
\item An embedding of \logic in the Rocq proof assistant with proofs of \emph{soundness} (no false alarms) and \emph{completeness}
(every incorrect execution admits a derivation),
the latter proved once and for all notions of incorrectness~\citep{anonymous-supplementary-material}.
\item Three instantiations of incorrectness:
stuckness (triggering undefined behavior),
the presence of a memory race
(two conflicting accesses, at least one non-atomic),
and the non-linearizability of a data structure (stated independently of any particular client).

\item An evaluation in which an LLM certifies seeded bugs with \logic, and finds and certifies two previously unknown bugs in the Saturn library~\citep{saturn}.
\end{itemize}

\logic has some limitations.
It builds on the ZooLang~\citep{allain-scherer-zoo-26} model of OCaml,
which does not model all of OCaml 5, so our completeness claims inherit its limitations.
In particular, by using a sequentially consistent model of OCaml 5, Mizzle is not able to certify bugs that depend on weak memory behaviors.
Furthermore, this paper does not claim to improve LLMs' ability to find bugs, nor do we argue that LLMs are a better method of finding bugs than other tools.
Rather, given that LLMs are being used for bug finding,
our goal with \logic{} is to provide a way to prove that an alleged bug is not a false alarm.

\section{Key Ideas}
\label{sec:key_ideas}
In this section, we present the key ideas of \logic.
We start with a motivating example:
a concrete buggy concurrent stack in \ocaml{}~(\cref{sec:key:example}),
and a companion test client in which only some interleavings
can trigger the bug.
To reason about this program, we
use \lang as a formal operational model of
\ocaml{}~(\cref{sec:key:zoo}).
\lang is not a contribution of this paper:
it is an almost direct copy of \zoolang~\citep{allain-scherer-zoo-26},
a model of OCaml 5 developed for reasoning about fine-grained concurrent programs in OCaml 5,
which we extend with support for atomic/non-atomic accesses.
Next, we present the key rules of \logic
and explain how to prove the existence of the bug by showing a stuck configuration is reachable~(\cref{sec:key:logic}).
\logic is in fact parameterized by the notion of incorrectness
being looked for, and stuckness is just one instantiation.
We illustrate this generality with a second instantiation:
non-linearizability, and how it can be applied to
our buggy concurrent stack~(\cref{sec:key:lin}).
We then state the soundness and completeness theorems of \logic,
specialized to both stuckness and non-linearizability~(\cref{sec:key:sound_complete}).

\subsection{A Buggy Concurrent Stack}
\label{sec:key:example}
\begin{figure}\centering
\begin{minipage}{0.58\textwidth}
\begin{minted}[fontsize=\small]{ocaml}
let create () = Atomic.make []

let rec push t v =
  let xs = Atomic.get t in
  let ys = v :: xs in
  (* Should be: Atomic.compare_and_set t xs ys *)
  if not (Atomic.compare_and_set t (Atomic.get t) ys)
  then push t v
\end{minted}
\end{minipage}
\hfill
\begin{minipage}{0.4\textwidth}
\begin{minted}[fontsize=\small]{ocaml}
let rec pop t =
  match Atomic.get t with
  | [] -> None
  | hd :: tl as xs ->
    if Atomic.compare_and_set t xs tl
    then Some hd else pop t
\end{minted}
\end{minipage}

\medskip
\begin{minipage}{0.58\textwidth}
\begin{minted}[fontsize=\small]{ocaml}
let client () =
  let t = create () in
  let f i = push t i; assert (pop t <> None) in
  ignore (Domain.spawn (fun () -> f 1));
  f 2
\end{minted}
\end{minipage}
\captionlabel{A buggy concurrent stack}{fig:buggy-stack}
\end{figure}

While \logic's ideas are language-agnostic,
we illustrate them on \ocaml{}, which natively supports shared-memory
concurrency via coarse-grained threads called \emph{domains}~\citep{sivaramakrishnan-20}.

\Cref{fig:buggy-stack}
presents an implementation of a (buggy) variant of Treiber's concurrent stack~\citep{treiber-86}.
The API of the stack is composed of three functions:
\ocamli{create ()}, which creates a new stack,
\ocamli{push t v}, which pushes a value \ocamli{v} on the stack \ocamli{t},
and \ocamli{pop t}, which pops a value from the stack~\ocamli{t},
wrapped in an option type to capture the case of an empty stack.
The stack is implemented as a reference
to an immutable list of values.
The function \ocamli{create ()} returns a reference to an empty list.
The function \ocamli{push t v} is buggy:
it loads the content of the stack \ocamli{t} in a local variable \ocamli{xs},
then builds a new list \ocamli{ys} by consing \ocamli{v} to \ocamli{xs}.
It then attempts to atomically install \ocamli{ys} with a compare-and-set~(CAS),
retrying on failure.
As we will see, the bug is that the CAS tests against a \emph{fresh} read of \ocamli{t}
instead of the snapshot \ocamli{xs}.
The function \ocamli{pop t} is correct:
it reads the stack and, if non-empty,
atomically replaces it with its tail using a CAS, retrying on failure.

\Cref{fig:buggy-stack} then presents a client which might
trigger the bug.
Two threads, one created with \ocamli{Domain.spawn} (returning a handle for a possible later join, which we ignore here),
push a value and then pop one.
Since each thread pops only after its own push,
each pop should return \ocamli{Some} of a value.
Yet, some interleavings of our client yield a pop operation
returning \ocamli{None}, which is a symptom of the buggy \ocamli{push}.
Concretely, the bug arises when both pushes take their snapshot \ocamli{xs}
of the empty stack before either commits.
The first CAS installs a single-element list.
The second then runs against this updated stack.
Because it compares the content of the stack against
a \emph{fresh} read of \ocamli{t}, with the single-element list,
the CAS succeeds and silently overwrites the first push.
The stack thus holds a single element instead of two,
so the second \ocamli{pop} finds it empty, returns \ocamli{None},
and fails the assertion.
This buggy behavior does not arise on every execution of the test client, as it requires a particular interleaving of the CAS operations.
For example, on an AMD Ryzen 7 PRO 250 with 8 cores, running Ubuntu 26.04 with OCaml 5.4.1,
it often takes more than a thousand runs to trigger the assertion failure.

\subsection{MizzleLang as a Model of \ocaml{}}
\label{sec:key:zoo}
In order to formally reason about \ocaml{} programs,
we need a \emph{model} of the language, that is,
a formal description of its syntax and semantics.
We base \logic on \lang,
which is almost a copy of \zoolang,
a recently developed model of \ocaml{} allowing for the verification of fine-grained concurrent \ocaml{}
algorithms with the Zoo program logic~\citep{allain-scherer-zoo-26}.
\lang captures many of the pertinent details of the
concurrency semantics of \ocaml{},
for example
including thread-local storage
and a subtle definition of physical equality.
Indeed, physical equality is underspecified in \ocaml{}
for values involving immutable blocks,
as sharing or other optimizations may render
two apparently equal values physically different,
and apparently different values physically equal.
As \citet{allain-scherer-zoo-26} mention,
this is a problem that can lead to bugs when using concurrency
primitives such as CAS, which test for physical equality.
In this paper, we assume sequential consistency.
The memory races we will later present~(\cref{sec:study:race})
all happen without weak memory considerations.

We present more of the formal details of \lang in \Cref{sec:syntax} and \Cref{sec:semantics}.
For now, it suffices to know that
we write a single step with the reduction relation
$\step{\config_1}{\config_2}$,
where $\config_1$ and $\config_2$ are two
\emph{configurations}, recording all the threads
currently executing, the heap, and a ghost trace of certain actions.
We write the reflexive-transitive closure
of the reduction relation as $\steprtc{\config_1}{\config_2}$.

\subsection{Mizzle at a Glance}
\label{sec:key:logic}
\logic is an \emph{angelic} logic:
it allows the user to prove the existence of one specific interleaving exhibiting a bug.
A \logic proof stops as soon as it reaches a witness of incorrectness chosen by the user~(\cref{sec:logic:goal}).
The rules of \logic follow the operational semantics of \lang:
the user can step any thread at any point,
which makes the logic complete~(\cref{sec:key:sound_complete}).
\logic is built using the Iris separation logic framework~\citep{iris}, and we follow its syntax for separation logic assertions.
We denote an Iris assertion by $\pre$,
a separating conjunction by $\pre \star \pre'$,
and a separating implication (also known as magic wand) by $\pre \wand \pre'$.
A meta-logical proposition $U$ is called \emph{pure} and written~$\pure{U}$.
The \emph{points-to assertion}
$\cells\loc{\vec\val}$ says that the memory location $\loc$
is the first location of a memory block that contains the values $\vec\val$,
that is, for any offset $\ofs$ such that $0 \leq \ofs < \sizeof{\vec\val}$,
the location $\loc + \ofs$ contains the value $\vec\val(\ofs)$.
\logic is not meant to be used by hand, but rather by an automated tool,
and in particular an LLM.
For simplicity, we will say ``the user''
throughout the paper
to refer to the agent using \logic.

To show that $\expr_c$, the \lang translation of the client expression in \Cref{fig:buggy-stack}, has a bug,
the user must prove the following entailment in Mizzle, which features the logic's three key assertions:
\[ \thread{\tid_0}{\nilctx}{\expr_c} \;\star\; \remaining{(\witness\stuckname)} \;\vdash\; \goal \]
Let us present the three assertions in this entailment from left to right.
First, the assertion $\thread{\tid}{\ectxs}{\expr}$ says
that a thread with identifier $\tid$ is executing the expression $\expr$
under evaluation context $\ectxs$.
In the above entailment,
$\tid$ is instantiated with $\tid_0$, the initial identifier,
$\ectxs$ with $\epsilon$, the empty evaluation context,
and $\expr$ with $\expr_c$, the client expression.
Second, the assertion $\remaining{\pre}$ states
the proof obligation that the user must show about the program's execution: the user has to exhibit $\pre$ to complete the proof.
In the above entailment,
$\pre$ is instantiated with $(\witness\stuckname)$,
where $\stuckname$
is a predicate over configurations
asserting that some thread is stuck,
that is, there exists a thread that has not reached
a value but cannot take a step.
In particular, any configuration with an \ocamli{assert false} is stuck
(a semantics where a failing assertion stops the execution and reports an error would be handled similarly, as another predicate over configurations).

One might wonder: if this $\remaining\pre$ assertion describes what the user has to \emph{show},
why is it on the left side of the entailment and not on the right?
The answer is that Mizzle uses a continuation-passing style, with
the $\goal$ assertion on the right side of the entailment as an opaque placeholder.
During a \logic proof, the right side of the entailment is always this $\goal$ assertion,
while the context holds one or more $\threadname$ assertions.
To step a thread, a rule consumes its $\threadname$ assertion
and asks the user to re-establish $\goal$ under an updated thread,
with the $\remainingname$ assertion telling us what we need to show to
discharge $\goal$.

\paragraph{Key reasoning rules}
\begin{figure}
\newcommand{\rulesep}{1.5ex}
\centering\small\morespacingaroundstar\morespacingaroundvdash
\begin{mathpar}
\inferrule[M-Assert]{}
{\thread{\tid}{\ectxs}{(\eassert{\vtrue})} \star (\thread{\tid}{\ectxs}{\vunit} \wand \goal) \vdash \goal}

\inferrule[M-CAS-Succ]{}
{\left(\,\begin{array}[c]{@{}r@{}}
\thread\tid\ectxs{(\ecas{(\loc, \ofs)}{\val_1}{\val_2})} \star{}\arcr
\cells\loc{\vec\val} \star{} \pure{0 \le \ofs < \sizeof{\vec\val} \;\wedge\; (\vread{\vec\val}\ofs \approx \val_1)} \star {}\arcr
(\cells\loc{\mapupd{\ofs}{\val_2}{\vec\val}} \star \thread\tid\ectxs{\vtrue} \wand \goal)
\end{array}\,\right) \vdash \goal}

\inferrule[M-CAS-Fail]{}
{\left(\,\begin{array}[c]{@{}r@{}}
\thread\tid\ectxs{(\ecas{(\loc, \ofs)}{\val_1}{\val_2})} \star{}\arcr\cells\loc{\vec\val}
  \star \pure{0 \le \ofs < \sizeof{\vec\val} \;\wedge\; (\vread{\vec\val}\ofs \napprox \val_1)} \star {}\arcr
(\cells\loc{\vec\val} \star \thread\tid\ectxs{\vfalse} \wand \goal)
\end{array}\,\right) \vdash \goal}

\mbox{$\inferrule[M-Fork]{}
{\left(\,\begin{array}[c]{@{}c@{}}
\thread\tid\ectxs{(\efork\expr)} \star {}\arcr
(\forall \tid'.\; \thread{\tid'}{\nilctx}\expr \star \thread\tid\ectxs\vunit \wand \goal)
\end{array}\,\right) \vdash \goal}
\quad\;\,
\inferrule[AssertStuck]{}{\thread{\tid}{\ectxs}{(\eassert{\val})} \star \pure{\val \neq \vtrue} \vdash \witness\stuckname}$}

\inferrule[RemainingUpdate]{}{\remaining{\pre_1} \star (\pre_2 \wand \pre_1) \vdash \remaining{\pre_2}}

\inferrule[M-Conclude]{}{\remaining{\iTrue} \vdash \goal}
\end{mathpar}
\captionlabel{Key reasoning rules of \logic: the assert function, the remaining assertion, forks, and compare-and-set}{fig:exrules}
\end{figure}

\cref{fig:exrules} presents a selection of reasoning rules of \logic, allowing us to verify the above entailment.
The first four rules of the figure are reasoning rules for advancing the execution of a thread.
\RULE{M-Assert} handles the case
where the assertion succeeds:
some thread~$\tid$ is executing $\eassert{\vtrue}$
under evaluation context $\ectxs$,
and the conclusion is $\goal$.
Then, the user can proceed
with the execution by assuming that
the thread $\tid$ faces the unit value $\vunit$ under evaluation context $\ectxs$,
and $\goal$ remains the target.
\RULE{M-CAS-Succ} and \RULE{M-CAS-Fail}
handle a compare-and-set $\ecas{(\loc, \ofs)}{\val_1}{\val_2}$,
which attempts to replace the value stored at location $\loc + \ofs$
by $\val_2$,
and succeeds only if that value is currently the expected value $\val_1$.
They both require the points-to assertion $\cells\loc{\vec\val}$
and that the offset $\ofs$ is in bounds.
The value currently stored at the location is thus $\vread{\vec\val}\ofs$.
The two rules compare it against the expected value $\val_1$
using physical equality~(\cref{sec:key:zoo}).
A succeeding comparison is captured by the relation
$\approx$ and a failed one by the relation $\napprox$.
Note that, because of the underspecification of physical equality in \ocaml{},
the two relations $\approx$ and $\napprox$ are not complementary.
On success~$(\vread{\vec\val}\ofs \approx \val_1)$,
the block is updated by storing $\val_2$ at offset $\ofs$;
on failure~$(\vread{\vec\val}\ofs \napprox \val_1)$, the block is left unchanged.
\RULE{M-Fork} spawns a new thread. The premise produces a fresh, distinct
identifier~$\tid'$ running $\expr$ under the empty evaluation context
$\nilctx$, while the forking thread~$\tid$ resumes with $\vunit$.

Although stated in continuation-passing style,
these rules are \emph{local} in the usual separation logic sense (that is, with a \emph{small footprint}):
each one mentions only the thread it steps and the memory it accesses.
The rest of the context is framed, as the following \RULE{Apply} rule shows:
\begin{mathpar}\morespacingaroundstar
\inferrule*[Left=Apply]
{\pre_1 \star (\pre_1' \wand \goal) \vdash \goal \\
 \pre_1' \star \pre_2 \vdash \goal}
{\pre_1 \star \pre_2 \vdash \goal}
\end{mathpar}
To make use of a lemma consuming $\pre_1$ and producing $\pre_1'$
while the context contains $\pre_1 \star \pre_2$,
the user applies \RULE{Apply}, and continues with $\pre_1' \star \pre_2 \vdash \goal$.
The resources $\pre_2$, such as other threads or memory locations,
are left untouched and passed to the continuation.

The last three rules of \Cref{fig:exrules} are reasoning rules for concluding a proof.
\RULE{AssertStuck} is a stuckness rule:
it allows for concluding $\witness\stuckname$
if some thread $\tid$ is executing $\eassert{\val}$ under evaluation context $\ectxs$,
and $\val$ is a value distinct from the Boolean $\vtrue$.
\RULE{RemainingUpdate} allows for replacing the obligation of a $\remainingname$ assertion
by a sufficient one, in particular for discharging it.
For example, the user can use this rule to
deduce $\remaining{(\witness\stuckname)} \star \witness\stuckname \vdash \remaining{\top}$,
where $\top$ is the always-true assertion.
\RULE{M-Conclude} is the sole rule of \logic
allowing one to finish the proof: it consumes $\remaining{\top}$,
the witness that no proof obligation remains,
and finally produces the target $\goal$.

\paragraph{Proving that our stack is buggy}
\bgroup\morespacingaroundstar\morespacingaroundvdash
\newcommand{\lnil}{\texttt{Nil}\xspace}
\newcommand{\lcons}{\texttt{Cons}\xspace}
We now sketch how these rules let us prove that the client of
\Cref{fig:buggy-stack} can get stuck.
For clarity, in this section,
we write in plain letters list constructors \lnil and \lcons.
Our target is to establish the entailment presented earlier:
\[\thread{\tid_0}{\nilctx}{\expr_c} \star \remaining{(\witness\stuckname)} \vdash \goal\]
Allocating the stack yields a points-to assertion $\cells\loc{[\lnil]}$
for a one-cell block whose single value is the empty list,
and \RULE{M-Fork} spawns the second thread, so that two threads
are both about to push, leading to the following entailment:
\[\thread{\tid_0}{\ectxs_a}{(\mathsf{push}\,\loc\,2)} \star
\thread{\tid_1}{\ectxs_a}{(\mathsf{push}\,\loc\,1)} \star
\cells\loc{[\lnil]} \star \remaining{(\witness\stuckname)}
\vdash \goal \]
where $\ectxs_a$ captures the pending $\mathsf{assert}$ for both threads.
The user is free to advance the threads in any order
by applying the rules from \Cref{fig:logic}, choosing the
interleaving that exhibits the bug.
We chose to make both threads enter the body of \ocamli{push},
and load $\loc$~(using \RULE{M-Load} from \Cref{fig:logic}).
They observe the \lnil constructor,
and build their respective new lists---that is,
the list \texttt{(Cons 2 Nil)} for thread~$\tid_0$,
and the list \texttt{(Cons 1 Nil)} for thread~$\tid_1$.
We now focus on thread $\tid_1$.
It performs a load on $\loc$ again, reading again the empty list,
and now faces a CAS operation:
\[\begin{array}{c}
\thread{\tid_0}{\ectxs_a}{(\mathsf{push}\,\loc\,2)} \star{}\\
\thread{\tid_1}{\ectxs_i}{(\mathsf{Atomic.compare\_and\_set}\,\loc\,\lnil\,\texttt{(Cons 1 Nil)})} \star{} \\
\cells\loc{[\lnil]} \star \remaining{(\witness\stuckname)}
\end{array} \vdash \goal \]
where $\ectxs_i$ captures the pending ``$\textsf{if not}$'' (which will fail)
and the follow-up \textsf{assert}.
Its CAS succeeds by \RULE{M-CAS-Succ},
storing \texttt{(Cons 1 Nil)} in $\loc$.
We then make~$\tid_1$ return from push and advance to its \textsf{assert} operation.
We now focus on thread~$\tid_0$.
It performs a load on $\loc$, but this time loads \texttt{(Cons 1 Nil)},
and faces its own CAS operation.
\[\begin{array}{c}
\thread{\tid_0}{\ectxs_i}{(\mathsf{Atomic.compare\_and\_set}\,\loc\,\texttt{(Cons 1 Nil)}\,\texttt{(Cons 2 Nil)})} \star{}\\
\thread{\tid_1}{\nilctx}{(\textsf{assert}\,(\textsf{pop}\,\loc <> \texttt{None}))} \star{} \\
\cells\loc{[\texttt{Cons 1 Nil}]} \star \remaining{(\witness\stuckname)}
\end{array} \vdash \goal \]
Here again,
because the buggy code compares against this \emph{fresh} read of $\loc$
rather than its previous snapshot, \RULE{M-CAS-Succ} applies,
and $\loc$ is overwritten.
The stack thus holds a single value \texttt{(Cons 2 Nil)}.
After both pushes return, thread~$\tid_0$ runs \ocamli{pop t},
which reads the singleton, succeeds with its CAS,
and returns \ocamli{Some}; its assertion holds by \RULE{M-Assert}.
Thread~$\tid_1$ then runs \ocamli{pop t} on the now-empty stack,
which returns \ocamli{None}, so its assertion reduces to $\eassert{\vfalse}$.
Now, the target is:
\[\thread{\tid_0}{\nilctx}{()} \star
\thread{\tid_1}{\nilctx}{(\textsf{assert}\,\vfalse)} \star{}
\cells\loc{[\lnil]} \star \remaining{(\witness\stuckname)}
\vdash \goal \]
We can now use \RULE{AssertStuck} to deduce $\witness\stuckname$,
and \RULE{RemainingUpdate} to consume the $\remaining{(\witness\stuckname)}$
obligation, leaving $\remaining{\top}$.
A final application of \RULE{M-Conclude} produces the target $\goal$,
completing the proof that the client can get stuck.
\egroup

\paragraph{Why use a logic instead of reasoning directly over the semantics?}
Since the primitive rules of \logic ``replay'' the steps of a buggy trace,
one might wonder what the logic offers over applying the operational semantics
to prove $\steprtc{\config_1}{\config_2}$ for the violating trace.
The answer is locality.
A step of the semantics rewrites a whole configuration:
every thread, the whole heap, and the trace appear in every step,
and nothing hides the parts a step does not touch.
This does not scale beyond toy programs,
and it forces the user to step through the implementation of every component involved.
Instead, \logic rules mention only the thread they
step and the memory they access;
the rest is framed.
Hence calls to library functions can be reasoned about with re-usable specifications rather than their code:
we provide a standard library for \lang
(arrays, dynamic arrays, atomic locations, domains, mutexes, condition variables, \etc)
whose functions come with \logic specifications.
Moreover, a specification lets \logic certify bugs under every behavior it permits:
our memory race case studies use an axiomatized specification of \parfor{}
that abstracts over how the parallelism in the iterations is chunked~(\cref{sec:study:race}).
Finally, since a \logic proof is a Rocq statement, a single proof can certify a bug for every input satisfying a predicate,
\eg, for all array sizes $n \geq 2$.

\subsection{Proving Non-Linearizability with \logic}
\label{sec:key:lin}

In order to show that the stack of \Cref{fig:buggy-stack}
is buggy, we had to exhibit a particular client.
Yet,
the defect lies not in the client, but in the stack itself:
it is \emph{not} linearizable.
Linearizability~\citep{herlihy-wing-90}
is a key correctness criterion for concurrent data structures.
Intuitively,
an operation on a data structure is linearizable
if, even though it may execute multiple computation steps,
its effect on the data structure appears to take place
instantaneously at a single point between its call and return, the so-called \emph{linearization point}.
Alternatively,
a history of concurrent operations on a data structure
is linearizable if it can be explained as a sequential history.
Non-linearizability constitutes the real issue with our running example:
the failing assertion of \Cref{sec:key:example} is only a symptom,
exhibited by one particular client.
Indeed, the stack admits histories
that no sequential stack could produce.

Linearizability is a \emph{trace} property, as one needs to consider the trace of function calls and returns
to determine whether a history is linearizable.
Thus, to expose the trace, we make use of a primitive $\eemit{\val}$,
which emits the value $\val$ to the trace, a piece of ghost state.

\begin{figure}
\centering\small\morespacingaroundstar\morespacingaroundvdash
\[\begin{array}{@{}r@{\;\;}c@{\;\;}l@{}}
\text{Model} & \Model & =\;\; \big(\Sset,\ \sinit \in \Sset,\ \valreq{} \subseteq \Values,\ \reqresp{}{}{}{} \subseteq \Sset \times \Values \times \Values \times \Sset\big)\\
\text{Event} & \eta & \eqdef\;\; \evcall{\val} \mid \evlin{\val}{\val'} \mid \evret{\val'} \\
\text{History} & \trace & \in\;\; \textsf{List}(\mathit{Tag} \times \text{Event}) \\
\end{array}\]
\begin{mathpar}
\inferrule[M-Emit]{}
{\begin{array}[c]{@{}r@{}}
\thread\tid\ectxs{(\eemit\val)} \star \istrace\trace \star {}\arcr
(\istrace{(\trace \concat [\val])} \star \thread\tid\ectxs\vunit \wand \goal)
\end{array} \vdash \goal}

\inferrule[NonLin]{}
{\begin{array}{@{}c@{}}
\pure{\nonlin\trace} \star{}\arcr \istrace\trace
\end{array}\vdash \witness\nonlinname}
\end{mathpar}
\captionlabel{Reasoning rules for non-linearizable traces}{fig:key:lin}
\end{figure}

We give the formal definition of linearizability later in \Cref{sec:defining_incorrectness}.
For now, \Cref{fig:key:lin} presents the relevant aspects needed to understand how a user proves non-linearizability.
First, we consider a \emph{model}~$\Model$ of the data structure, which is a specification of how a sequential version of
the data structure, such as a stack or queue, behaves.
A model is a tuple of a set of states~$\Sset$,
an initial state~$\sinit$,
a set~\valreq{} of valid arguments for the operation,
and a relation~$\reqresp{s}{\val}{\val'}{s'}$
describing the permissible transitions from state~$s$ to state~$s'$ when a call with argument~$\val$ returns~$\val'$.
We then define the notion of an \emph{event}~$\eta$:
either a call with some argument~$\val$,
a linearization of a call with argument~$\val$ returning~$\val'$,
or the effective return point of a call, returning~$\val'$.
We modify operations to emit the call and return events.
Meanwhile, the linearization event is not emitted; instead, given a trace, the definition of linearizability
requires annotating
the trace with linearization events in a way that is consistent with some sequential behavior of the model.
A \emph{history}~$\trace$ is a list of tagged events,
where the tag, a natural number, allows for matching up call events to their corresponding returns in the history.
We write that a history $\trace$ is \emph{non-linearizable}
as $\nonlin{\trace}$,
when all its calls are done with valid arguments
and yet it cannot be explained by any sequential history~(\Cref{sec:defining_incorrectness}).

For dealing with traces, \logic provides the assertion $\istrace{\trace}$,
witnessing that the observation trace is currently $\trace$.
The lower part of \Cref{fig:key:lin} presents two reasoning rules of \logic,
key for proving non-linearizability using traces.
First, \RULE{M-Emit} appends a value to the trace.
It requires the assertion
$\istrace\trace$, and returns it updated to
$\istrace{(\trace \concat [\val])}$, the thread reducing to
the unit value~$\vunit$.
Second, \RULE{NonLin} allows for concluding the assertion
$(\witness\nonlinname)$,
if we can deduce the pure meta-logic fact that the current trace $\trace$ is non-linearizable.

Putting these pieces together,
to prove that a data structure is not linearizable,
it suffices to exhibit an execution that generates a non-linearizable trace,
that is, a succession of calls and returns that cannot be explained as a sequential history.

\paragraph{The most general client}
\begin{figure}
\settowidth{\mgcwidth}{\small\ttfamily\ \ \ \ (ignore (Domain.spawn (fun () -> onecall i)); client (i + 1)) in}
\centering
\begin{minipage}{\mgcwidth}
\begin{minted}[fontsize=\small]{ocaml}
let most_general_client init op gen =
  let obj = init () in
  let onecall i =
    let x = gen () in emit (i, Call x);
    let r = op obj x in emit (i, Ret r) in
  let rec client i =
    if nondet_bool () then () else
    (ignore (Domain.spawn (fun () -> onecall i)); client (i + 1)) in
  client 1
\end{minted}
\end{minipage}
\captionlabel{The most general client}{fig:mgc}
\end{figure}

We cannot just consider any possible client that might interact with the data structure: a client that violates the abstraction boundaries of the data structure and directly modifies its internal state is not a demonstration of a bug in the data structure.
We avoid this issue of invalid clients and relieve the user from having to construct clients explicitly by defining
a function called the \emph{most general client}~(a standard technique, see \citet{dongol-derrick-15}), presented in \Cref{fig:mgc}.
For simplicity, we suppose that the data structure
is equipped with a single operation taking a single argument and returning a value.
Data structures with multiple operations can be emulated by passing a tuple with a tag to indicate which operation to select.

The function \ocamli{most_general_client} takes three arguments:
\ocamli{init}, initializing the data structure,
\ocamli{op}, applying an operation on the data structure,
and \ocamli{gen}, generating arguments for the operation.
A \ocamli{most_general_client} call
initializes the data structure, then repeatedly nondeterministically spawns threads.
Each thread runs one function call,
emitting a \ocamli{Call} event before the call
and a \ocamli{Ret} event after it.
We call this the most general client because, for every execution of a valid client of the data structure,
there is some choice of \ocamli{gen} and a corresponding execution of the \ocamli{most_general_client}
that yields the same interleaving of calls.

\paragraph{Proving non-linearizability for our buggy stack}
We instantiate the most general client with our stack,
taking \ocamli{init} to be \ocamli{create}
and letting \ocamli{op} dispatch on a tagged argument to invoke either \ocamli{push} or \ocamli{pop}.
We use a \ocamli{gen} function that randomly generates either pops or pushes with an argument of \ocamli{1} or \ocamli{2}.
We then use the rules of Mizzle to generate an execution that produces the trace
\[\trace \;\eqdef\;
\begin{array}[t]{@{}l@{}}
[\,
\tagged{1}{\evcall{(\mathsf{push}\,1)}};\,
\tagged{2}{\evcall{(\mathsf{push}\,2)}};\,
\tagged{1}{\evret{\vunit}};\,
\tagged{2}{\evret{\vunit}};\,\\
\phantom{[\,}
\tagged{3}{\evcall{\mathsf{pop}}};\,
\tagged{3}{\evret{(\mathsf{Some}\,2)}};\,
\tagged{4}{\evcall{\mathsf{pop}}};\,
\tagged{4}{\evret{\mathsf{None}}}\,]
\end{array}\]
where the two pushes overlap, as required by the lost update,
and both return before either pop is called.
This history is not linearizable:
since the two pops follow both completed pushes,
any sequential explanation would have to pop the two pushed values,
returning $\mathsf{Some}$ twice.
Yet, operation~$4$ observes $\mathsf{None}$,
the symptom of the lost push,
and $\nonlin\trace$ holds.
In particular, in our Rocq mechanization, since the definitions of the sequential stack model are computable, the user can apply a decision procedure that shows that no linearization of $\trace$ exists.

The proof that there exists an interleaving of the call
to the most general client exhibiting the above trace reuses
the
interleaving of \Cref{sec:key:logic}.
Let us write $\expr_{mgc}$ for the expression that invokes the most general client
with the arguments described above.
The initial entailment to prove is:
\[\morespacingaroundstar \thread{\tid_0}{\nilctx}{\expr_{mgc}} \star \istrace{\epsilon} \star \remaining{(\witness\nonlinname)} \vdash \goal\]
With helper lemmas provided by \logic
(one for starting the most general client,
one for initiating a call, and one for returning from a call),
we then drive the four threads so as to produce~$\trace$.
Each \ocamli{emit} is justified by \RULE{M-Emit}:
it consumes the current $\istrace{\trace'}$ and returns
$\istrace{(\trace' \concat [\val])}$,
appending one event at a time until the trace is exactly $\trace$.
Once the full history has been emitted,
\RULE{NonLin} applies: it requires $\istrace\trace$ and the pure fact
$\nonlin\trace$ established above,
and produces
$(\witness\nonlinname)$, witnessing non-linearizability.
This discharges the remaining proof obligation,
so \RULE{RemainingUpdate} leaves $\remaining\top$
and \RULE{M-Conclude} produces the target $\goal$.

\subsection{Soundness and Completeness}
\label{sec:key:sound_complete}
\logic is sound
(intuitively,
if the user proves a $\goal$ assertion,
then they have effectively constructed a buggy execution)
and complete
(intuitively, if there exists a buggy execution, there exists a \logic derivation).
\Cref{sec:soundness,sec:completeness} present
a general soundness theorem and
a general completeness theorem,
parameterized by the notion of ``incorrectness'' being considered.
In this section we state corollaries of these theorems
for two of our instantiations, stuckness and non-linearizability.

\paragraph{Soundness}
The soundness theorem says that a \logic proof
with a $\remaining (\witness\pre)$ proof obligation
effectively constructs a reduction path
starting from $\confinit{\expr}$,
the initial configuration with an empty heap and empty trace,
that witnesses the incorrectness property $\pre$.
For stuckness, the constructed reduction path reaches a stuck configuration,
whereas for non-linearizability, the reduction path instead emits a non-linearizable history.
\begin{restatable}[Soundness for Stuckness]{corollary}{corsoundnessstuck}
\label{thm:cor:soundness_stuck}
Let $\expr$ be an expression and $\tid_0$ the initial thread identifier.
If $\thread{\tid_0}{\nilctx}{\expr} \star \remaining{(\witness\stuckname)} \vdash \goal$
holds,
then there exists a configuration $\config$
such that $\steprtc{\confinit{\expr}}{\config}$ and $\stuck\config$ hold.
\end{restatable}
\begin{restatable}[Soundness for Non-Linearizability]{corollary}{corsoundnessnonlin}
\label{thm:cor:soundness_nonlin}
Let $\expr$ be a call to the most general client and $\tid_0$ the initial thread identifier.
If $\thread{\tid_0}{\nilctx}{\expr} \star \istrace{\epsilon} \star \remaining{(\witness\nonlinname)} \vdash \goal$
holds,
then there exists a configuration $\config$
such that $\steprtc{\confinit{\expr}}{\config}$ and $\nonlin\config$ hold.
\end{restatable}

\paragraph{Completeness}
Conversely, the completeness theorem
assumes the existence of a reduction path from
the initial configuration $\confinit{\expr}$
reaching an incorrect configuration $\config$,
and concludes that there exists a \logic derivation of $\goal$.
The one caveat is that it assumes that $\expr$ contains no hard-coded locations,
because \logic has no rules to reason about them.
This is not a serious restriction, since a valid \ocaml{} program that is translated with the \texttt{ocaml2zoo} tool into \lang{}
cannot have such locations, because well-typed \ocaml{} programs do not provide a way to refer directly to the memory address backing
a reference.
\begin{restatable}[Completeness for Stuckness]{corollary}{corcompletenessstuck}
\label{thm:cor:completeness_stuck}
Let $\expr$ be an expression.
If $\locs{\expr} = \emptyset$ and $\steprtc{\confinit{\expr}}{\config}$ holds
with $\stuck\config$,
then $\thread{\tid_0}{\nilctx}{\expr} \star \remaining{(\witness\stuckname)} \vdash \goal$
holds.
\end{restatable}
\begin{restatable}[Completeness for Non-Linearizability]{corollary}{corcompletenessnonlin}
\label{thm:cor:completeness_nonlin}
Let $\expr$ be a call to the most general client.
If $\locs{\expr} = \emptyset$ and $\steprtc{\confinit{\expr}}{\config}$ holds
with $\config$ having a location-free, non-linearizable history,
then $\thread{\tid_0}{\nilctx}{\expr} \star \istrace{\epsilon} \star \remaining{(\witness\nonlinname)} \vdash \goal$
holds.
\end{restatable}

Note that the corollary for non-linearizability
assumes that the emitted history contains no locations.
This precondition is a simplification:
as we explain in \Cref{sec:completeness},
the general completeness theorem makes strict assumptions
about locations because \logic treats them abstractly.

\section{Syntax and Semantics}
\label{sec:syntax_and_semantics}
In this section, we briefly describe \lang.
\lang and its semantics are not a contribution of this paper,
as they are an almost direct copy of \zoolang~\citep{allain-scherer-zoo-26}.
\lang adds support for distinguishing atomic from non-atomic
accesses, which is needed to faithfully model a memory race~(\cref{paragraph:racy}).
\lang drops support for prophecies, which are not needed in the context of an incorrectness
logic, and instead adds an $\eemit$ primitive to record observations in a trace;
such a trace is later used to define the notion of non-linearizability~(\cref{paragraph:nonlin}).
Finally, \lang makes \oob accesses stuck,
instead of letting them silently read or write a neighboring block;
we explain this choice in \appref{sec:appendix:oob}.

We first describe the syntax~(\cref{sec:syntax}) and then
the operational semantics~(\cref{sec:semantics}).

\subsection{Syntax}
\label{sec:syntax}
\begin{figure}
\centering\small
\newcommand{\commentary}[1]{ & \text{\small\it #1} \\}
\[
\begin{array}{l@{\quad}r@{\;\eqdef\;}l}
\text{Atomicity} & \atm & \matomic \mid \mnonatomic
  \qquad\qquad \text{Unary ops}\quad \unop \eqdef {\sim} \mid -
  \qquad\qquad \text{Binary ops}\quad \binop \eqdef + \mid \times \mid \ldots \\

\text{Values}\;\Values & \val& \vbool \in \Bool \mid \vint \in \Int \mid \loc \in \Loc \mid \vrec\funcname\argname\body \\

\text{Expressions} & \expr &
\begin{array}[t]{@{}l@{\hspace{8mm}}l@{}}
\begin{array}[t]{@{}ll@{}}
\val,\wal
\commentary{value}
\var
\commentary{variable}
\erec\funcname\argname\expr
\commentary{abstraction}
\eapp\expr\expr
\commentary{call}
\elet\var\expr\expr
\commentary{let binding}
\eunop\unop\expr
\commentary{unary operation}
\ebinop\expr\binop\expr
\commentary{binary operation}
\eequal\expr\expr
\commentary{physical equality}
\eif\expr\expr\expr
\commentary{conditional}
\end{array} &
\begin{array}[t]{@{}ll@{}}
\eblock\tagn{\vec{\expr}}
\commentary{block construction}
\ealloc\expr\expr
\commentary{array allocation}
\egettag\expr
\commentary{block tag}
\egetsize\expr
\commentary{block size}
\eload\expr\expr
\commentary{load}
\estore\expr\expr\expr
\commentary{store}
\ecas\expr\expr\expr
\commentary{compare-and-set}
\efork\expr
\commentary{fork}
\eemit\expr
\commentary{trace emission}
\end{array}
\end{array}\\

\text{Contexts} & \ectx &
\begin{array}[t]{@{}llll}
\eapp\khole\val &\mid \eapp\expr\khole &\mid \elet\var\khole\expr &\mid \eunop\unop\khole \\
\mid \eif\khole\expr\expr &\mid \ebinop\khole\binop\val &\mid \ebinop\expr\binop\khole &\mid \eequal\khole\val \\
\mid \eequal\expr\khole &\mid \eblock\tagn{(\vec\expr \concat \khole \concat \vec\val)} &\mid \egettag\khole &\mid \egetsize\khole \\
\mid \eload\khole\val &\mid \eload\expr\khole &\mid \estore\khole\val\val &\mid \estore\expr\khole\val \\
\mid \estore\expr\expr\khole &\mid \ecas\khole\val\val &\mid \ecas\expr\khole\val &\mid \ecas\expr\expr\khole \\
\mid \ealloc\khole\val &\mid \ealloc\expr\khole &\mid \eemit\khole
\end{array}
\end{array}\]
\captionlabel{Syntax of \lang}{fig:syntax}
\end{figure}

\Cref{fig:syntax} presents the syntax of \lang, a call-by-value
lambda calculus with mutable state and concurrency
modeling \ocaml{}'s semantics.
A value~$\val \in \Values$ is either a Boolean~$\vbool \in \Bool$,
an idealized integer~$\vint \in \Int$, a location~$\loc$ from an infinite
set of locations~$\Loc$, or a recursive function~$\vrec\funcname\argname\body$.
We also write $\vunit$ for the unit value; in the mechanization, it is the empty immutable tuple~(\appref{sec:appendix:differences}).
A \emph{block} is a heap-allocated record, denoted by its location~$\loc$,
carrying a constructor tag~$\tagn \in \mathbb{N}$ and a sequence of fields~$\vec{\val}$.
Constructor tags allow distinguishing between different constructors of an algebraic data type.
For example, in the case of the type \ocamli{option} in OCaml,
the data \ocamli{None} is represented by a block with tag~$0$ and no field,
while the data \ocamli{Some v} is represented by a block with tag~$1$ and the single field~$v$.
(The Rocq implementation offers notations and machinery to
manipulate OCaml's algebraic constructors instead of raw tags and blocks.)

A computation is described by an expression~$\expr$, whose syntax is
mostly standard. Unary and binary primitive operations
$\eunop\unop\expr$ and $\ebinop\expr\binop\expr$ cover the usual
arithmetic, bitwise and comparison operators, while $\eequal\expr\expr$
tests physical equality.
Blocks model both records and the constructors of algebraic data types:
$\eblock\tagn{\vec{\expr}}$ allocates a block with tag~$\tagn$ and
fields~$\vec{\expr}$.
The expressions $\egettag\expr$ and $\egetsize\expr$ retrieve a block's tag
and number of fields.
The expression $\ealloc\tagn m$ allocates a fresh block of $m$ fields
(all initialized to $\vunit$), where $m$ may be a runtime value.
The expression
$\eload\expr\expr$ reads a field at a given offset of a block,
$\estore\expr\expr\expr$ writes it, and the atomic primitive
$\ecas\expr\expr\expr$ provides compare-and-set,
where the first argument is a tuple of a location and an offset.
Loads and stores carry an atomicity flag~$\atm$, either atomic ($\matomic$) or non-atomic ($\mnonatomic$).
This flag does not influence the interleaving semantics of \lang,
but it determines which accesses take part in a memory race~(\cref{paragraph:racy}).
The primitive $\eemit\val$ appends the value~$\val$ to a global observation
trace, and returns the unit value~$\vunit$.
Finally, $\efork\expr$ spawns a concurrent thread.
A (non-recursive) evaluation context~$\ectx$ describes an
expression with a hole~$\khole$
and dictates the right-to-left evaluation order of \lang.
We write $\ectx[\expr]$ for the expression obtained by filling the hole of~$\ectx$ with~$\expr$,
and write similarly $\ectxs[\expr]$ when multiple contexts are stacked.

The previous \Cref{sec:key_ideas} features the $\eassert{x}$ function,
which is defined in \lang as
$\eassert{x} \eqdef \eif{x}{\vunit}{\vunit\,\vunit}$,
the function that returns the unit value $\vunit$
if its argument is $\vtrue$ and becomes stuck otherwise,
by applying the unit value to itself.

\paragraph{Differences with the mechanization}
For simplicity, this paper presents a subset of \lang.
Our mechanization
supports the full language, including immutable structures,
pattern matching, loops,
additional atomic operations, and thread-local state.
Our soundness \Cref{thm:soundness}
and completeness \Cref{thm:completeness}
are verified against the full semantics.
We list the differences in \appref{sec:appendix:differences}.

\subsection{Semantics}
\label{sec:semantics}
\begin{figure}\centering\small\morespacingaroundstar

\begin{mathpar}
\inferrule[Alloc]
  {0 \leq m \\ \disjoint{\big(\{\loc\} \cup \{\loc + \ofs \mid 0 \leq \ofs < m\}\big)}{\big(\dom(\tagmap) \cup \dom(\store)\big)}}
  {\headstep{\ealloc\tagn m}{\tagmap}{\store}{\trace}{\loc}{\mapupd{\loc}{\headerval{\tagn}{m}}{\tagmap}}{\mapupd{\loc + \ofs}{\vunit \mid 0 \leq \ofs < m}{\store}}{\trace}{\epsilon}}
\\
\inferrule[Load]
  {\store(\loc + \ofs) = \val \\ \tagmap(\loc) = \headerval{\tagn}{m} \\ 0 \leq \ofs < m}
  {\headstep{\eload\loc\ofs}{\tagmap}{\store}{\trace}{\val}{\tagmap}{\store}{\trace}{\epsilon}}

\inferrule[Store]
  {\loc + \ofs \in \dom(\store) \\ \tagmap(\loc) = \headerval{\tagn}{m} \\ 0 \leq \ofs < m}
  {\headstep{\estore\loc\ofs\val}{\tagmap}{\store}{\trace}{\vunit}{\tagmap}{\mapupd{\loc + \ofs}{\val}{\store}}{\trace}{\epsilon}}
\\
\inferrule[Fork]{}
  {\headstep{\efork\expr}{\tagmap}{\store}{\trace}{\vunit}{\tagmap}{\store}{\trace}{[\expr]}}

\inferrule[Emit]{}
  {\headstep{\eemit\val}{\tagmap}{\store}{\trace}{\vunit}{\tagmap}{\store}{\trace \concat [\val]}{\epsilon}}
\\
\inferrule[Context]
  {\headstep{\expr}{\tagmap}{\store}{\trace}{\expr'}{\tagmap'}{\store'}{\trace'}{\exprs_f}}
  {\threadstep{\fillk{\vec\ectx}\expr}{\tagmap}{\store}{\trace}{\fillk{\vec\ectx}{\expr'}}{\tagmap'}{\store'}{\trace'}{\exprs_f}}

\inferrule[Interleave]
  {\exprs(\tid) = \expr \\
   \threadstep\expr\tagmap\store\trace{\expr'}{\tagmap'}{\store'}{\trace'}{\exprs_f}}
  {\step{\mkconf{\exprs}{\store}{\tagmap}{\trace}}{\mkconf{\mapupd{\tid}{\expr'}{\exprs} \concat \exprs_f}{\store'}{\tagmap'}{\trace'}}}
\end{mathpar}
\captionlabel{Selected rules of the semantics of \lang}{fig:semantics}
\end{figure}

A \emph{configuration} $\config = \mkconf{\exprs}{\store}{\tagmap}{\trace}$
is a quadruple of a list of expressions $\exprs$, one per thread, a heap $\store$, a tag map~$\tagmap$, and a trace $\trace$.
The heap $\store$ is a partial function mapping locations to values.
The tag map~$\tagmap$
records, for each allocated block,
its header $\headerval{\tagn}{m}$: its constructor tag $n$ and its size~$m$.
The trace $\trace$ is the list of values emitted so far.

\Cref{fig:semantics} presents selected rules of the semantics of \lang,
we give further rules, for block construction and compare-and-set, in \appref{sec:appendix:semantics}.
The \emph{head reduction relation} $\headstep{\expr}{\tagmap}{\store}{\trace}{\expr'}{\tagmap'}{\store'}{\trace'}{\exprs_f}$
describes a single computation step of the expression~$\expr$.
It relates the expression and the
state, including the trace $\trace$, to their reducts,
together with the
(possibly empty, written~$\epsilon$)
list $\exprs_f$ of threads spawned by the step.
For example,
\RULE{Alloc} allocates a block of $m$ fields, all initialized to~$\vunit$,
at a location~$\loc$ chosen nondeterministically among the fresh ones,
that is, such that neither $\loc$ nor its $m$ cells are in the domain of the tag map or of the heap.
It records the header $\headerval{\tagn}{m}$ in the tag map;
in particular, if $m = 0$, the block has a header but no cell.
\RULE{Load} and \RULE{Store} respectively read and write the field at offset~$\ofs$
of the block at location~$\loc$.
Both use the header of the block to check that $\ofs$ is within bounds,
and are stuck otherwise.
Their atomicity flag~$\atm$ plays no role in the semantics.
\RULE{Fork} spawns a thread and reduces to the unit value.
\RULE{Emit} reduces to the unit value and appends its argument~$\val$ to
the trace, yielding $\trace \concat [\val]$.
The \emph{deep reduction relation} $\threadstep{\expr}{\tagmap}{\store}{\trace}{\expr'}{\tagmap'}{\store'}{\trace'}{\exprs_f}$
closes head reduction under evaluation contexts:
\RULE{Context} locates the next redex deep inside~$\expr$, following the right-to-left
evaluation order dictated by the contexts~$\ectx$ of \cref{fig:syntax},
and executes one head step.
The \emph{reduction relation} $\step{\config_1}{\config_2}$ operates on
whole configurations.
\RULE{Interleave} nondeterministically
selects one thread~$\tid$ from the pool~$\exprs$, takes a deep step on it, and
updates the configuration accordingly,
replacing thread~$\tid$ by its reduct,
appending any spawned threads~$\exprs_f$, and carrying over the updated
trace~$\trace'$. This last relation models the
interleaving concurrency of OCaml~5's domains.

\section{The Mizzle Program Logic}
\label{sec:logic}
In this section, we present \logic in depth.
We first give its reasoning rules for the cases when ``things go well'',
that is, when no incorrectness is witnessed~(\cref{sec:rules_well}).
We then explain how the user defines their own notion of incorrectness,
and illustrate it with stuckness, memory races, and non-linearizability~(\cref{sec:defining_incorrectness}).
We next state and discuss the general
soundness~(\cref{sec:soundness}) and completeness~(\cref{sec:completeness}) theorems.
Finally, we take a closer look at the definition of the $\goal$ assertion~(\cref{sec:logic:goal}).

\subsection{Reasoning Rules When Things Go Well}
\label{sec:rules_well}
\begin{figure}
\newcommand{\rulesep}{2.5ex}
\centering\small\morespacingaroundstar\morespacingaroundvdash
\[\begin{array}{@{}c@{\quad}r@{\;\vdash\;}l@{}}
\TirNameStyle{M-Block} & \mlprec{
\thread\tid\ectxs{(\eblock\tagn{\vec\val})} \star \pure{0 < \sizeof{\vec\val}} \star {}\\
(\forall \loc.\; \header\loc\tagn{\sizeof{\vec\val}} \star \cells\loc{\vec\val} \star \thread\tid\ectxs\loc \wand \goal)
}
  & \goal \\[1.5ex]

\TirNameStyle{M-Alloc} & \mlprec{
\thread\tid\ectxs{(\ealloc\tagn m)} \star \pure{0 \le m} \star {}\\
(\forall \loc.\; \header\loc\tagn{m} \star \cells\loc{{\vunit}^{m}} \star (\pure{m = 0} \wand \loctok\loc) \star \thread\tid\ectxs\loc \wand \goal)
}
  & \goal \\[\rulesep]

\TirNameStyle{M-Load} & \mlprec{
\thread\tid\ectxs{(\eload\loc\ofs)} \star \cells\loc{\vec\val}
  \star \pure{0 \le \ofs < \sizeof{\vec\val}} \star {}\\
(\cells\loc{\vec\val} \star \thread\tid\ectxs{(\vread{\vec\val}\ofs)} \wand \goal)
}
  & \goal \\[\rulesep]

\TirNameStyle{M-Store} & \mlprec{
\thread\tid\ectxs{(\estore\loc\ofs\wal)} \star \cells\loc{\vec\val}
  \star \pure{0 \le \ofs < \sizeof{\vec\val}} \star {}\\
(\cells\loc{\mapupd{\ofs}{\wal}{\vec\val}} \star \thread\tid\ectxs\vunit \wand \goal)
}
  & \goal\\[\rulesep]
\TirNameStyle{M-Context} & \multicolumn{2}{r}{\prowstrut\thread\tid{(\ectxs_1 \concat \ectx_2)}\expr \;\dashv\!\vdash \thread\tid{\ectxs_1}{(\fillk{\ectx_2}\expr)}}\\
\end{array}\]
\captionlabel{Reasoning rules for block construction, allocation, loads, and stores}{fig:logic}
\end{figure}

To represent mutable state,
\logic features two assertions.
First, the points-to assertion, as described in \Cref{sec:key:logic}.
Second, a companion assertion $\header\loc\tagn{m}$ witnesses that the memory location
$\loc$ contains a block of size $m$ tagged with $\tagn$.
This assertion is \emph{persistent},
hence in particular duplicable.

\Cref{fig:logic} presents \logic's reasoning rules
when no form of incorrectness is witnessed.
In fact, each of these rules corresponds to a step of the
incorrectness derivation:
each premise mentions a $\thread\tid\ectxs\expr$,
some resources to execute the step faced by $\expr$
and a continuation-style assertion that describes the resources and threads after the step.

\RULE{M-Block} allocates a block from the values $\vec\val$,
provided the block is non-empty (${0 < \sizeof{\vec\val}}$). In exchange,
the user obtains, for a freshly chosen location $\loc$, the header assertion
$\header\loc\tagn{\sizeof{\vec\val}}$, the points-to
$\cells\loc{\vec\val}$, and the thread now reduced to the value $\loc$.
\RULE{M-Alloc} is the analogue for the $\ealloc\tagn m$ primitive: it
allocates a fresh block of $m$ fields, all initialized to $\vunit$.
Unlike \RULE{M-Block}, the size $m$ is an arbitrary runtime value
subject only to $0 \le m$, so the block may be empty. In exchange, the
user obtains, for a freshly chosen location $\loc$, the header assertion
$\header\loc\tagn{m}$, the points-to ${\cells\loc{{\vunit}^{m}}}$, and the
thread reduced to $\loc$. The additional, conditional assertion
$\pure{m = 0} \wand \loctok\loc$ hands back the live-location token
$\loctok\loc$ only in the degenerate case $m = 0$.
This token allows the user to separate a block with size 0 from another location,
with the rule $\loctok{\loc_1} \star \loc_2 \mapsto \vec\val \star \pure{0 < |\vec\val|} \vdash \pure{\loc_1 \neq \loc_2}$.
\RULE{M-Load} reads the location $\loc + \ofs$.
It requires the assertion $\cells\loc{\vec\val}$ and an
in-bounds index ($0 \leq \ofs < \sizeof{\vec\val}$);
the points-to assertion is
returned unchanged, and the thread reduces to $\vread{\vec\val}\ofs$.
\RULE{M-Store} writes the value $\wal$ into location $\loc + \ofs$.
It requires the assertion $\cells\loc{\vec\val}$ and returns it updated to
$\cells\loc{\mapupd{\ofs}{\wal}{\vec\val}}$, the thread reducing to $\vunit$.
\RULE{M-Context} allows for rearranging evaluation contexts,
from the second parameter of the $\threadname$
assertion to the expression itself.
Note that \logic has no \textsc{Bind} rule,
that is, a rule allowing for forgetting
about the evaluation context while focusing
on the current expression.

\subsection{Defining Incorrectness}
\label{sec:defining_incorrectness}
In \logic, the user can define their own notion of incorrectness,
or more precisely, the property they want to hold on
execution traces constructed by the user.
We saw two examples in \Cref{sec:key_ideas}: stuckness and non-linearizability.
This section gives more formal detail on how these notions of incorrectness are defined,
and defines a third notion, memory races.
We first present a key assertion:
the $\witnessname$ assertion, together with
the state interpretation predicate.

\paragraph{Demanding a witness with the state interpretation predicate}
\begin{figure}
\centering\small\morespacingaroundstar\morespacingaroundvdash
\[\begin{array}{@{}c@{\qquad}r@{\;\vdash\;}l@{}}
\TirNameStyle{InterpThread} & \interp{(\mkconf{\vec\expr}{\store}{\tagmap}{\trace})} \star \thread\tid\ectxs{\expr} & \pure{\vec\expr(\tid) = \fillk{\ectxs}{\expr}} \\
\TirNameStyle{InterpPointsTo} & \interp{(\mkconf{\vec\expr}{\store}{\tagmap}{\trace})} \star \cells\loc{\vec\val}
  & \pure{\forall \ofs.\; 0 \le \ofs < \sizeof{\vec\val} \Rightarrow \store(\loc + \ofs) = \vread{\vec\val}\ofs} \\
\TirNameStyle{InterpHeader} & \interp{(\mkconf{\vec\expr}{\store}{\tagmap}{\trace})} \star \header\loc\tagn{m} & \pure{\tagmap(\loc) = \headerval\tagn{m}} \\
\TirNameStyle{InterpTrace} & \interp{(\mkconf{\vec\expr}{\store}{\tagmap}{\trace})} \star \istrace{\trace'} & \pure{\trace' = \trace}
\end{array}\]
\captionlabel{Reasoning rules for the \interpname predicate}{fig:interp}
\end{figure}

In order to relate separation logic assertions to
the state of the operational semantics,
we make available to the user the
\emph{state interpretation predicate},
written $\interp\config$,
where $\config = \mkconf{\exprs}{\store}{\tagmap}{\trace}$ is a configuration
with threads $\exprs$, heap~$\store$, tag map $\tagmap$, and trace $\trace$.
Intuitively, if $\interp\config$ holds,
it means that the current execution trace constructed by the user
reached the configuration $\config$.
Formally, $\interp\config$ is a standard Iris state interpretation,
relating $\config$ to the ghost state defining the $\threadname$, points-to, $\headername$, and trace assertions.

\Cref{fig:interp}
presents reasoning rules for exploiting
the \threadname, points-to, header, and trace assertions,
assuming that $\interp{(\mkconf{\exprs}{\store}{\tagmap}{\trace})}$ holds.
\RULE{InterpThread} allows for deducing that
if $\thread\tid\ectxs{\expr}$ holds,
then $\fillk\ectxs\expr$ is the $\tid$-th expression of $\exprs$.
\RULE{InterpPointsTo} allows for deducing that
if $\cells\loc{\vec\val}$ holds,
then for any valid offset $\ofs$,
$\store(\loc + \ofs)$ is the $\ofs$-th value of $\vec\val$.
\RULE{InterpHeader} allows for deducing that
if $\header\loc\tagn{m}$ holds,
then $\tagmap(\loc)$ is the header $\headerval\tagn{m}$.
\RULE{InterpTrace} allows for deducing that
if $\istrace{\trace'}$ holds,
then $\trace'$ is the configuration's trace $\trace$.

We then define the assertion \witnessname as:
\[\witness{P} \;\eqdef\; \forall\config.\; \interp\config \wand \pure{P\,\config}\]
Intuitively, if $\witness{P}$ holds,
it means that under the assumption that the current execution trace reached a configuration $\config$,
the property $P\,\config$ holds.

\paragraph{Stuckness}
\begin{figure}
\centering\small\morespacingaroundstar\morespacingaroundvdash
\[\stuck{{(\mkconf{\vec\expr}{\store}{\tagmap}{\trace})}} \eqdef \exists \tid\,\expr_\tid.\;\exprs(\tid) = \expr_\tid \wedge \expr_\tid \notin \Val \wedge \neg (\exists \expr'\,\store'\,\tagmap'\,\trace'\,\exprs_f.\; \threadstep{\expr_\tid}{\tagmap}{\store}{\trace}{\expr'}{\tagmap'}{\store'}{\trace'}{\exprs_f})\]
\[\begin{array}{@{}c@{\qquad}r@{\;\vdash\;}l@{}}
\TirNameStyle{StuckLoad} & \thread\tid\ectxs{(\eload{\val_1}{\val_2})} \star
  \pure{\val_1 \notin \Loc \lor \val_2 \notin \Int}
  & \witness\stuckname \\

\TirNameStyle{StuckLoadOob} & \header\loc\tagn{m} \star \thread\tid\ectxs{(\eload\loc\ofs)} \star \pure{\ofs < 0 \lor m \le \ofs}
  & \witness\stuckname
\end{array}\]
\captionlabel{The \stuckname{} predicate, and stuckness rules for load operations}{fig:incstuck}
\end{figure}

\Cref{fig:incstuck}
presents the definition of the $\stuckname$ predicate and rules for proving that a thread is stuck during a load operation.
The predicate $\stuck\config$ holds
if there exists a thread $\tid$ whose expression $\expr_\tid$
is not a value, and that cannot take a step in the semantics.
The rules for load cover two cases.
\RULE{StuckLoad} captures the ill-typed case, that is,
if the load operation is not applied to a location,
or if the offset is not an integer.
\RULE{StuckLoadOob} captures the out-of-bound case.
The out-of-boundness is exhibited by the header assertion.
From the definition of $(\witness\stuckname)$,
the reader might infer how we prove, for example, \RULE{StuckLoadOob}.
We use the header to deduce that the offset is wrong using \RULE{InterpHeader},
then use the $\threadname$ assertion to deduce that indeed
a load operation is being executed on this offset using \RULE{InterpThread},
and conclude.

\paragraph{Memory races}
\label{paragraph:racy}
\begin{figure}
\centering\small\morespacingaroundstar\morespacingaroundvdash
\begin{mathpar}
\mprset{andskip=1em plus 0.5fil minus 0.5em}
\inferrule[LAccLoad]{}
{\locacc{(\eload\loc\ofs)}{(\loc + \ofs)}{\bot}{\atm}}

\inferrule[LAccStore]{}
{\locacc{(\estore\loc\ofs\wal)}{(\loc + \ofs)}{\top}{\atm}}

\inferrule[LAccCAS]{}
{\locacc{(\ecas{\vtuple{\loc, \ofs}}{\val}{\wal})}{(\loc + \ofs)}{\top}{\matomic}}

\inferrule[AccCtx]
{\locacc{\expr}{\loc}{b}{\atm}}
{\cacc{\fillk\ectxs\expr}{\loc}{b}{\atm}}
\end{mathpar}
\[\begin{array}{@{}r@{\;}c@{\;}l@{}}
\racy{(\mkconf{\vec\expr}{\store}{\tagmap}{\trace})} & \eqdef & \exists \tid_1\,\expr_1\,\tid_2\,\expr_2\,\loc\,b_1\,b_2\,\atm_1\,\atm_2.\; \vec{\expr}(\tid_1) = \expr_1 \wedge \vec{\expr}(\tid_2) = \expr_2 \wedge \tid_1 \neq \tid_2 \wedge {} \\
&& \cacc{\expr_1}{\loc}{b_1}{\atm_1} \wedge \cacc{\expr_2}{\loc}{b_2}{\atm_2} \wedge (b_1 \lor b_2) \wedge (\mnonatomic = \atm_1 \lor \mnonatomic = \atm_2)
\end{array}\]
\vspace{-0.5em}
\[\begin{array}{@{}c@{\quad}c@{\;\vdash\;}c@{}}
\TirNameStyle{IsRacy} &
\begin{array}[c]{@{}c@{}}
\thread{\tid_1}{\ectxs_1}{\expr_1} \star \thread{\tid_2}{\ectxs_2}{\expr_2}\star {} \\
\pure{\exists \loc\,b_1\,b_2\,\atm_1\,\atm_2.\; \locacc{\expr_1}{\loc}{b_1}{\atm_1} \wedge \locacc{\expr_2}{\loc}{b_2}{\atm_2} \wedge (b_1 \lor b_2) \wedge (\mnonatomic = \atm_1 \lor \mnonatomic = \atm_2)}
\end{array}
& \isracy
\end{array}\]
\captionlabel{The \locaccname and \caccname access predicates, and the \racyname{} predicate}{fig:incdetrace}
\end{figure}

Another important class of bugs consists of \emph{memory races}:
two concurrent accesses to a shared memory location,
with at least one of them being a write operation
and at least one of them being non-atomic.
We define a predicate $\racyname$ which identifies configurations
where two threads are concurrently accessing the same location,
at least one of them writing and at least one of them non-atomically---%
recall that loads and stores of \lang carry an atomicity flag
(which is ignored by the interleaving semantics),
while compare-and-set is always considered atomic~(\cref{sec:syntax}).
Although \lang is sequentially consistent,
a race certified by \logic is a race of the \ocaml{} program:
every sequentially consistent execution is allowed by the \ocaml{} memory model,
and two accesses that are both poised in a configuration are unordered
by the happens-before relation.
In our evaluation~(\cref{sec:study:race}), all accesses
but the ones in the implementation of synchronization primitives such as locks
are non-atomic.
Synchronization requires no further machinery,
in particular no trace of past accesses.
For example, if two threads protect their accesses
to a location with the same lock,
then no reachable configuration has both threads
concurrently accessing this location.
We detail one implementation of locks and their specification in \appref{sec:appendix:lock}.

In our setting, a memory race manifests itself purely syntactically
on the configuration.
The upper part of
\Cref{fig:incdetrace} presents the $\locacc{\expr}{\loc}{b}{\atm}$
predicate, capturing that expression $\expr$ accesses location~$\loc$;
the boolean $b$ indicates whether the access may be a write operation,
and $\atm$ is the atomicity of the access.
The predicate $\cacc{\expr}{\loc}{b}{\atm}$ lifts the $\locaccname$ predicate
to arbitrary evaluation contexts.
\Cref{fig:incdetrace} then presents the
property $\racy{(\mkconf{\vec\expr}{\store}{\tagmap}{\trace})}$,
formalizing a memory race:
it asserts the existence of two different threads which are
accessing the same location, with at least one of them being
a write operation, and at least one of them being non-atomic.

The lower part of \Cref{fig:incdetrace}
presents \RULE{IsRacy},
the main reasoning rule allowing the user to prove
the $(\isracy)$ separation logic assertion.
The rule consumes two threads $\tid_1$ and $\tid_2$ executing expressions $\expr_1$ and $\expr_2$,
each one of them satisfying the $\locaccname$ predicate on the same location $\loc$,
with at least one of the two accesses being a write operation,
and at least one of them being non-atomic.
In order to prove \RULE{IsRacy},
we unfold the definition of $\witnessname$,
and get that $(\isracy)$ is equivalent to
$\forall\config.\; \interp\config \wand \pure{\racy{\config}}$.
Hence, we use the $\threadname$ assertion to deduce that indeed
two threads are executing $\expr_1$ and $\expr_2$
under some evaluation contexts,
then use the separating conjunction to deduce that $\tid_1 \neq \tid_2$,
and conclude.

\paragraph{Non-linearizability}
\label{paragraph:nonlin}
\begin{figure}
\centering\small\morespacingaroundstar
\begin{mathpar}
\inferrule[SoundLinRefl]{}{\drives{s}{\epsilon}{s}}

\inferrule[SoundLinCons]{\drives{s_0}{\trace}{s} \\ \valreq{\val} \\ \reqresp{s}{\val}{\val'}{s'}}
          {\drives{s_0}{(\trace \concat [\tagged{\optag}{\evlin{\val}{\val'}}])}{s'}}
\end{mathpar}

\[\begin{array}{@{}r@{\;\;\defeq\;\;}l@{}}
\callvalid{\trace} & \forall \ofs\,\optag\,\val.\; \trace(\ofs) = (\optag,\evcall\val) \implies \valreq{\val} \\
\pertaglin{\trace} & \forall \optag.\; \exists \val\,\val'.\
  \proj{\trace}{\optag} \;\prefixeq\;
  [\,\tagged{\optag}{\evcall{\val}};\, \tagged{\optag}{\evlin{\val}{\val'}};\, \tagged{\optag}{\evret{\val'}}\,] \\
\linof{\trace'}{\trace} &
  \trace = (\delin{\trace'}) \;\wedge\; \pertaglin{\trace'} \;\wedge\; \exists s'.\ \drives{\sinit}{(\onlin{\trace'})}{s'}\\
\nonlin{\trace} &
  \callvalid{\trace} \;\wedge\; \neg (\exists \trace'.\ \linof{\trace'}{\trace})\\
\end{array}\]
\[\nonlin{(\mkconf{\exprs}{\store}{\tagmap}{\trace})} \;\eqdef\; \nonlin{\trace}\]
\captionlabel{Linearizability as a trace property}{fig:inclin}
\end{figure}

\Cref{fig:inclin} presents the formal definition of linearizability,
relying on the model~$\Model$ introduced in \Cref{fig:key:lin}.
As alluded to in \Cref{fig:key:lin}, the purpose of the linearization events is to annotate
the trace with markers indicating in what order the corresponding operations are applied.
The upper part of the figure presents the judgment
$\drives{s_0}{\trace}{s}$,
asserting that applying the linearization events of $\trace$ in the order they occur
updates the model from state~$s_0$ to state~$s$.
\RULE{SoundLinRefl} handles the empty trace, leaving the state unchanged.
\RULE{SoundLinCons} extends a trace with one linearization event
$\tagged{\optag}{\evlin{\val}{\val'}}$,
provided the argument~$\val$ is valid, and that the model admits the
transition $\reqresp{s}{\val}{\val'}{s'}$.
The lower part of \Cref{fig:inclin} then defines trace predicates.
First, $\callvalid{\trace}$ requires every call event in~$\trace$ to carry a valid argument.
Second, $\pertaglin{\trace}$ requires that, for each tag~$\optag$,
the projection $\proj{\trace}{\optag}$ of $\trace$,
selecting only the events associated with that tag,
is a prefix of a well-formed operation
$[\,\tagged{\optag}{\evcall{\val}};\, \tagged{\optag}{\evlin{\val}{\val'}};\, \tagged{\optag}{\evret{\val'}}\,]$;
that is, each operation calls, linearizes, then returns, in this order,
with matching arguments and results.
Third, $\linof{\trace'}{\trace}$ asserts that $\trace'$ is a linearization of $\trace$,
that is, $\trace'$ enriches $\trace$ with linearization events
and satisfies several properties.
The first property requires that $\trace$ is $\trace'$ deprived of its
linearization events, $\delin{\trace'}$.
Next, $\trace'$ is well-formed per tag,
meaning that for each tag $\tau$,
if we project $\trace'$ down to just the events with tag $\tau$, the projected trace is a prefix of a call/linearize/return sequence.
Finally, applying those linearization events must update the model from~$\sinit$ to some state~$s'$.
A trace~$\trace$ is then non-linearizable, written $\nonlin{\trace}$,
if its calls are valid and it admits \emph{no} linearization.
Likewise, a configuration is non-linearizable, written $\nonlin\config$, when its trace is.

\subsection{Soundness of \logic}
\label{sec:soundness}

The soundness theorem of \logic
states that if the user can prove $\goal$
from the initial thread~$\tid_0$ executing $\expr$
with a proof obligation $(\witness{P})$,
for some meta-level predicate $P$ on configurations,
then there exists a configuration $\config$ reachable from the initial configuration $\confinit{\expr}$
such that $P\,\config$ holds.
The initial configuration $\confinit{\expr}$ starts with the empty heap and the empty trace~$\epsilon$.
\begin{theorem}[Soundness of \logic]
\label{thm:soundness}
Let $\expr$ be an expression, $\tid_0$ the initial thread identifier,
and~$P$ a meta-level predicate on configurations.
If $\thread{\tid_0}{\nilctx}{\expr} \star \remaining{(\witness{P})} \vdash \goal$ holds,
then there exists a configuration~$\config$
such that $\steprtc{\confinit{\expr}}{\config}$ and $P\,\config$ hold.
\end{theorem}
Using \Cref{thm:soundness},
we prove
\Cref{thm:cor:soundness_stuck} for stuckness,
a similar theorem for memory races (omitted for conciseness),
and \Cref{thm:cor:soundness_nonlin} for non-linearizability,
by instantiating $P$ to $\stuckname$, $\racyname$, and $\nonlinname$, respectively.
Note that
\Cref{thm:cor:soundness_nonlin} specializes
$\expr$ to be a call to the most general client.
The statement holds without this restriction,
but guarantees nothing about the placement of emit operations.

\subsection{Completeness of \logic}
\label{sec:completeness}
\begin{figure}
\centering\small\morespacingaroundstar
\begin{align*}
\mirror{(\mkconf{\exprs}{\store}{\tagmap}{\trace})} \;\;\eqdef\;\;\exists&\,\mapping.\;
  \pure{\mirrorinvs{\mapping}{\exprs}{\store}{\tagmap}}
\star\bigast{(\tid \mapsto \expr)\,\in\,\exprs}{\left(\thread{\tid}{\nilctx}{\tr{\mapping}{\expr}}\right)}\\
&\star\bigast{(\loc \mapsto \headerval\tagn m)\,\in\,\tagmap}{\left(\header{\tr{\mapping}{\loc}}\tagn m \star \bigast{0 \le \ofs < m}{\left(\cells{(\tr{\mapping}{\loc} + \ofs)}{\tr{\mapping}{\store(\loc + \ofs)}}\right)}\right)}\\
&\star\bigast{\loc\,\in\,\dom\,\tagmap \setminus \dom\,\store}{\left(\loctok{\tr{\mapping}{\loc}}\right)}
\star\istrace{\tr{\mapping}{\trace}}
\end{align*}
\captionlabel{The \mirrorname assertion}{fig:mirror}
\end{figure}

Since the rules of \logic follow the operational semantics,
one may expect completeness to hold by construction:
follow the reduction path, applying one rule per step.
This is not the case,
because \logic abstracts over details of the execution that the user does not control.
\RULE{M-Block} and \RULE{M-Fork} return universally quantified locations and thread identifiers,
so a derivation reaches the configuration of the reduction only up to a renaming.
This renaming must preserve the block structure of the heap, so that offsets commute with it;
conversely, it cannot force two blocks to be adjacent,
so \lang makes out-of-bounds accesses stuck rather than potentially letting them reach a neighboring block~(\appref{sec:appendix:oob}).
The invariant relating reduction and derivation must moreover be preserved by every step,
including the allocation of empty blocks and the emission of trace events.
The $\mirrorname$ assertion of \cref{fig:mirror} is this invariant.
In the following, if $\mapping$ is a partial mapping from locations to locations,
we write $\tr{\mapping}{x}$ for the translation of $x$ under $\mapping$,
where $x$ is an expression or a value.

To state completeness properly, we need to restrict attention to incorrectness properties that do not depend on specific
choices of allocation locations.
\Cref{fig:mirror} presents the $\mirror{(\mkconf{\exprs}{\store}{\tagmap}{\trace})}$ assertion,
which formalizes this idea.
This assertion requires the existence of
a bijection $\mapping$ between the locations of the configuration
and the locations of the separation logic assertions.
The fact that~$\mapping$ is a bijection with the correct domain
is captured by the pure assertion $\mirrorinvs{\mapping}{\exprs}{\store}{\tagmap}$.
The $\mirrorname$ assertion then asserts,
for each thread $\tid$ executing $\expr$ in the configuration,
the ownership of the translated $\thread{\tid}{\nilctx}{\tr{\mapping}{\expr}}$.
Next, the $\mirrorname$ assertion
asserts the ownership of headers and points-to assertions for each allocated location,
whose values are adequately translated.
Then, the $\mirrorname$ assertion asserts ownership of
live-location tokens for each allocated location that is not in the domain of the heap---that is, of 0-sized blocks.
Finally, the $\mirrorname$ assertion asserts the ownership of the trace
assertion $\istrace{\tr{\mapping}{\trace}}$, capturing the configuration's
emitted trace, up to renaming by~$\mapping$.

We can then devote our attention to the completeness theorem of \logic.
Formally speaking, this theorem states
that, for a program $\expr$ with no hard-coded locations,
if there exists a configuration~$\config$ reachable from the initial configuration $\confinit{\expr}$,
then there exists a \logic derivation with a remaining proof obligation $\mirror\config$.
\begin{theorem}[Completeness of \logic]
\label{thm:completeness}
Let $\expr$ be an expression and $\config$ a configuration. If $\locs\expr = \emptyset$,
and if $\steprtc{\confinit{\expr}}{\config}$ holds,
then $\thread{\tid_0}{\nilctx}{\expr} \star \remaining{(\mirror{\config})} \vdash \goal$ holds.
\end{theorem}

\Cref{thm:completeness} is proved once, independently of any notion of incorrectness:
intuitively, $\mirror\config$ is the strongest assertion that can be proved about $\config$ in \logic.
Completeness for a given notion~$P$ then only requires a lemma $\mirror\config \vdash \witness P$
for the configurations~$\config$ satisfying~$P$,
after which \RULE{RemainingUpdate} concludes.

To derive the completeness \Cref{thm:cor:completeness_stuck}
for stuckness, we prove a lemma whose conclusion is $\mirror\config \vdash \witness{\stuckname}$.
Its $\stuck\config$ hypothesis is exactly
the hypothesis of \Cref{thm:cor:completeness_stuck},
while the other hypotheses
correspond to invariants of the reduction relation.
\begin{lemma}
If (1) $\stuck\config$ holds,
(2) every location mentioned in a thread has an associated header in the tag map,
and (3) the locations in the domain of the tag map are present in the heap,
then $\mirror\config \vdash \witness{\stuckname}$.
\end{lemma}

As for a completeness corollary for memory races,
which we do not state here for conciseness,
we propose a lemma whose conclusion is $\mirror\config \vdash \witness{\racyname}$.
The first hypothesis is exactly the one witnessing $\racy\config$,
while the in-bounds requirement on the two accesses
corresponds to an invariant of the reduction relation.
\begin{lemma}
If $\config$ has two distinct threads $\tid_1 \neq \tid_2$
executing expressions $\expr_1$ and $\expr_2$
such that $\locacc{\expr_1}{\loc}{b_1}{\atm_1}$ and $\locacc{\expr_2}{\loc}{b_2}{\atm_2}$ hold
for the same location $\loc$ with $b_1 \vee b_2$ and $\atm_1 = \mnonatomic \lor \atm_2 = \mnonatomic$,
and both accesses are in bounds,
then $\mirror\config \vdash \witness{\racyname}$.
\end{lemma}

To derive the completeness \Cref{thm:cor:completeness_nonlin}
for non-linearizability,
we prove a lemma whose conclusion is $\mirror\config \vdash \witness{\nonlinname}$.
Its hypotheses
are exactly the ones from \Cref{thm:cor:completeness_nonlin}.%
\begin{lemma}
If $\config$'s trace is non-linearizable and contains no locations,
then \\$\mirror\config \vdash \witness{\nonlinname}$ holds.
\end{lemma}
We can now explain the need for the trace to not include any locations.
Indeed, recall that the trace assertion held by $\mirror\config$ is
$\istrace{\tr{\mapping}{\trace}}$, capturing the configuration's trace
up to the renaming~$\mapping$.
When the trace contains no locations, the renaming acts as the identity,
so the trace held by $\mirror\config$ is \emph{equal} to
the configuration's trace, and non-linearizability transfers directly.

\subsection{Inside the Goal Assertion}
\label{sec:logic:goal}
\begin{figure}\centering\small\morespacingaroundstar
\begin{align*}
\goal \;\eqdef\;\; & \forall\config.\; \interp\config \wand \pvs{} (\interp\config \star \remaining\top) \,\lor\, (\exists\config'.\; \pure{\step\config{\config'}} \star \interp{\config'} \star \goal)
\end{align*}
\captionlabel{Definition of the $\goal$ assertion}{fig:goal}
\end{figure}

The definition of $\goal$
mirrors the proof of \Cref{thm:soundness}.
Indeed, we define $\goal$ as a \emph{least fixpoint},
allowing us to perform reasoning by induction.
Either the base case holds,
and the user has fulfilled their remaining proof obligation,
or there exists one possible step,
after which we can apply the induction hypothesis.
In the cases where we take a step, we need to reflect how that step updates the physical state of the system.
Recall from \Cref{sec:defining_incorrectness}
that the physical state and the logical state
are related using the state interpretation predicate $\interpname$.
Thus, as the program makes progress,
one needs to update logical assertions accordingly.
Such an update is possible thanks to the Iris \emph{ghost update modality},
written $\pvs{} \pre$, asserting that $\pre$ holds after such an update.

\Cref{fig:goal} presents
the definition of the $\goal$ assertion.
It generalizes the assertion of the same name of \citet{moine-westrick-tassarotti-26},
whose base case requires the execution to have terminated:
in \logic, the base case is reached as soon as the user discharges $\remaining\top$,
that is, the proof obligation they chose.
The definition universally quantifies
over the configurations $\config$,
and requires ownership of the state interpretation predicate $\interp\config$.
Then, it allows for a ghost update thanks to the ghost update modality $\pvs{}$.
Next, the definition consists of a disjunction of two cases.
Either $\interp\config$ holds untouched, together with the $\remaining\top$ assertion,
witnessing that the initial proof obligation has been fulfilled.
This is the base, non-recursive case.
Otherwise,
the definition asserts the existence of a configuration~$\config'$ reachable from $\config$ in one step.
It also returns the ownership of the updated state interpretation predicate $\interp{\config'}$,
as well as the $\goal$ assertion itself.

\section{Case Studies}
\label{sec:case_studies}
In this section, we (a) evaluate whether an LLM can in fact use \logic
to prove the existence of a bug, and (b)
assess the cost of constructing a formal proof that a bug exists.
We first present our methodology (\cref{sec:study:methodo}),
and then two case studies, where bugs were manually seeded:
memory races (\cref{sec:study:race}) and non-linearizability (\cref{sec:study:nonlin}).
We then explain how we tasked an agent
to look for potential non-linearizability
bugs and to verify them
in the open-source Saturn library~\citep{saturn} (\cref{sec:study:saturn}).
With this approach, we discovered and reported two previously unknown bugs.

\subsection{Methodology}
\label{sec:study:methodo}

The programs we consider in this section are all written in \ocaml and
translated into \lang using an adaptation of the \texttt{ocaml2zoo} tool~\citep{allain-scherer-zoo-26}.
For memory races, we ported a
number of examples from the \emph{DataRaceBench} suite~\citep{dataracebench},
which we describe in more detail in~\Cref{sec:study:race}.
For non-linearizability bugs, we adapted implementations of concurrent
data structures from the Saturn library~\citep{saturn} and manually inserted bugs.
We first confirmed that our mechanization had all the lemmas and tactics needed,
by working interactively with LLMs to construct a \emph{reference solution}
for a subset of our test cases.
Our case studies evaluate the cost both on this tuning set and on a held-out test set.

The tactic library provides one tactic per language primitive that applies the corresponding proof rule (\eg, one for loads, one for stores, \dots).
While we did initially provide tactics to execute multiple steps, we observed that the LLM systematically preferred to use per-primitive steps---perhaps because this gives more fine-grained control over how it reduces the program.
A key observation we make
is that an LLM is yet another user,
who needs good tactic support and description.

Based on our experience developing the reference solutions, we wrote one
\emph{\logic{} proof guide} per family of case studies (memory races
and non-linearizability), which we provide as a prompt to each agent,
together with a short description of the project architecture.
Both guides share the same structure.
Each opens with a warning that \lang evaluates right-to-left
(\ocaml famously leaves evaluation order unspecified, but in practice both the bytecode and native compilers evaluate arguments right-to-left),
a pitfall the model repeatedly ran into during interactive development.
It then gives:
(1)~which steps are performed by dedicated tactics,
and how to apply the remaining lemmas by hand;
(2)~a workflow for interacting with Rocq through the MCP server;
(3)~a complete proof of one representative case study;
(4)~an API map, recalling the notations of \lang{}
and associating each primitive with its tactic;
(5)~a family-specific section: how to step through loops using an invariant
for memory races, and how to design a witness trace for
non-linearizability; and
(6)~an API card, generated automatically from the Rocq sources,
listing the statements of the key lemmas.
We kept the guides and tactics as general as possible, and report results
not only on the ``tuning set'' used to develop them, but also on a
``held-out set'' of examples added after freezing them.
The provided reference solution
reduces the verification cost by about 10\% for memory race examples,
and by up to 40\% for non-linearizability examples---likely
because our methodology for proving non-linearizability
is less standard and more complex than the one for memory races.

In \Cref{sec:study:race,sec:study:nonlin}, we report the results of running
the model on each test case. For a given test case, we delete all other
reference solutions and task the model with proving the designated
top-level theorem, which it is forbidden from weakening or restating.

For these case studies, we used \emph{Claude Opus 5.5}.
The model interacts with Rocq through an MCP server~\citep{rocqmcp}.
We asked the model to ensure that no axioms were used,
which we then manually checked while confirming that the final proof state preserved the desired statement.
A \logic certificate is thus a closed Rocq proof:
it trusts only Rocq, \lang as a model of \ocaml,
the \texttt{ocaml2zoo} translation,
and the axiomatized specifications the proof explicitly uses.
We ran each test case three times.
In every run, the agent found and proved the bug.

\paragraph{Limitations and threats to validity}
Before discussing the results, let us first acknowledge several limitations of the experimental setup.
This experiment does \emph{not} evaluate how good LLMs are at finding unknown bugs, for several reasons:
(1) \emph{DataRaceBench} is widely used, so there may already have been leakage;
(2) the non-linearizability bugs are relatively simple errors that we inserted,
which may not be representative of real bugs.
As we stated in \Cref{sec:intro}, the goal of this paper is \emph{not} to assess or improve LLM bug finding.
Rather, the purpose of these experiments is to assess the cost of constructing a formal proof that a bug exists.

Aside from this, these experiments suffer from a \emph{framing} bias:
the model is told that the program is incorrect and asked to prove this,
rather than to assess if the program is incorrect or not before proving.
While we do not tell the agent where the bug is, this still potentially reduces the overall cost compared to having to determine first if the program has a bug or not.
However, we envision that a possible workflow in practice might be to use one agent to first find bug candidates, and then use a second agent to fully determine the bug and construct the incorrectness proofs. In that case, this setup is closer to the scenario the second agent runs in.

\subsection{Memory Races}
\label{sec:study:race}
We now evaluate the ability of an LLM to prove the presence of a
memory race.

\paragraph{Test cases}
The test cases we use
for this study are direct OCaml translations
of some examples of the \emph{DataRaceBench} (DRB) suite~\citep{dataracebench}.
DRB examples are written in C/C++ and use
the OpenMP library for parallelism (\eg, a parallel for loop).
We selected a subset of DRB examples that were easily translatable to
OCaml, using the \texttt{domainslib} library~\citep{domainslib} for
lightweight parallelism in a similar style as OpenMP.
We also selected examples
that were \emph{positive} (\ie, they truly had a bug; some entries have no bugs).
We refer to each case by its identifier in DataRaceBench
(\eg, case 005 is the port of \texttt{DRB005-indirectaccess1-orig-yes.c}).
DRB examples come with a ``test'', that is, a final assertion potentially failing due to a memory race.
We removed this test, in order to not give the model any hint about where the race is located.
(Our approach does not rely on this test to identify bugs, so it is safe to
remove.)
Each case study is between 4 and 44 OCaml lines of code,
and the model outputs between 23 and 129 lines of Rocq.

The \texttt{domainslib} library is complex. %
In order to avoid stepping into the library
using \logic,
we axiomatize some primitives~(an approach we discuss further in
\Cref{sec:conclusion}).
In particular, we provide the following axiom for the parallel for loop:
{\small\morespacingaroundstar
\[\begin{array}[c]{@{}r@{}}
\thread\tid\ectxs{(\parfor\,n\,\body)} \star {}\\
\Big(
\forall \vec\mu\, \vec\eta.\;
\left(\kern-1em\begin{array}{@{}c@{}}
\pure{\sizeof{\vec\mu} = \sizeof{\vec\eta} = n} \star (\bigsep_{i < n} \thread{\vec\mu(i)}{\vec\eta(i)}{(\body\,i)}) \star{}\\
\quad\big(
  (\bigsep_{i < n} \thread{\vec\mu(i)}{\vec\eta(i)}{\vunit}) \star
  (\thread\tid\ectxs\vunit \wand \goal) \wand \goal
  \big)
\end{array}\right)\wand\goal\Big)
\end{array}
\;\vdash\;\goal
\]}%
The axiom states that a thread $\tid$ about to run
$\parfor\,n\,\body$ under
evaluation context $\ectxs$
may spawn~$n$ iteration threads $\vec\mu$, each
running $\body\,i$ with its own thread identifier $\vec\mu(i)$
and under its own evaluation context $\vec\eta(i)$.
Once every iteration has been driven to $\vunit$, the user regains the parent
thread $\tid$, with $\parfor\,n\,\body$ reduced to $\vunit$, and must
re-establish $\goal$.

Note that this specification does not hold as-is
if the parallel for loop sequentially chunks some operations, as
the \texttt{domainslib} library does.
Yet, we argue that this specification is the one
that the user should use for bug finding, as the size of chunks is usually a performance configuration option:
the correctness of the program should \emph{not} depend on this parameter having a particular value.
This axiomatized specification allows us to search for bugs across all choices of this parameter (see case 179 below).
As a sanity check, we prove in \logic that a divide-and-conquer implementation of the parallel for loop,
written in \lang on top of domains, satisfies this specification.

\paragraph{Results}
\begin{figure}
\centering\small
\begin{tabular}{llrrrrr}
\toprule
\textbf{Set} & \textbf{Family} & \textbf{Cases} & \textbf{Input tokens} & \textbf{Output tokens} & \textbf{Cost} & \textbf{Time} \\ \midrule
\multirow{2}{*}{Tuning} & Shared cell & 5 & 262K & 5K & \$0.35 & 50s \\
 & Loop-carried & 7 & 333K & 7K & \$0.47 & 71s \\
\midrule
\multirow{3}{*}{Held-out} & Shared cell & 4 & 241K & 6K & \$0.38 & 61s \\
 & Loop-carried & 8 & 415K & 9K & \$0.53 & 88s \\
 & Missing sync. & 10 & 307K & 6K & \$0.40 & 61s \\
\bottomrule
\end{tabular}
\captionlabel{Evaluation of \logic being used by Claude Opus 5.5 to prove the existence of a memory race. %
The numbers are averaged over the cases of each family and over 3 runs. %
Between runs, the cost of a case varies by $\pm5\%$ on average (at most $\pm21\%$). %
Per case, the cost ranges from \$0.23 to \$1.00 (average \$0.43).}{fig:drb_res:summary}
\end{figure}

\Cref{fig:drb_res:summary} presents the results.
We classify memory races into three categories.
First, shared cell bugs,
that is, parallel for loops accessing a shared cell without proper synchronization.
Second, loop-carried bugs, that is,
when a particular iteration $i$ of a parallel for loop
depends on a write operation made \emph{by another, concurrent iteration}, for example $i+1$.
Third, missing synchronization bugs,
which involve mis-handled locking and unlocking operations.

The cost is low: \$0.43 per example on average, and at most \$1.00.
The missing synchronization family appears only in the held-out set.
Its programs rely on locks from a standard library,
which come with specifications that the proof guide never mentions.
A distinctive feature of \logic is that such specifications can be proved
rather than assumed, since we can implement locks in \lang and reason about them,
see \Appref{sec:appendix:lock}.

\paragraph{A comparison point with dynamic race detection}
As a check of our translation,
we ran each port under ThreadSanitizer~\citep{serebryany-iskhodzhanov-09}.
We used OCaml 5.5.1 with 5 domains,
and ran each port 100 times.
ThreadSanitizer reports a race in the benchmark code for 31 of the 34 cases,
confirming independently that these ports are racy.
Some races are rarely observed (cases 005 and 006 are reported in under 5\% of the runs).
ThreadSanitizer reports no race in the three remaining cases.
The first one (case 189) diverges due to a mishandling of locks.
The second one (case 185) has a window for the race
so short that the main domain completes its part before the second domain has finished spawning.
The third one (case 179)
has a race consisting of iterations 0 and 1
of a parallel for loop
writing to a common cell,
but with the default chunk size of \texttt{domainslib},
these two iterations belong to the same chunk and run sequentially.

\subsection{Non-Linearizability}
\label{sec:study:nonlin}

In \Cref{sec:key:lin},
we explained how \logic can be used to prove
that a concurrent data structure is non-linearizable.
In this section,
we test this approach on a set of case studies.
They all correspond to standard concurrent data structures
in which we manually inserted bugs.
The tuning set contains
Treiber's stack~\citep{treiber-86},
Michael and Scott's lock-free queue~\citep{michael-scott-96},
and a concurrent hash table~\citep{saturn}.
The held-out set contains
Michael and Scott's two-lock queue~\citep{michael-scott-96}
and Harris's lock-free linked list~\citep{harris-01}.
We asked an agent to prove non-linearizability
following our methodology (\cref{sec:study:methodo}).
Note that, for each data structure,
we provide the sequential specification of the data structure
the agent is expected to use in its proof,
as explained in \Cref{sec:key:lin}.
Each case study is between 24 and 62 OCaml lines of code,
and the model outputs between 81 and 156 lines of Rocq.
\begin{figure}
\centering\small
\begin{tabular}{llrrrrr}
\toprule
\textbf{Set} & \textbf{Structure} & \textbf{Cases} & \textbf{Input tokens} & \textbf{Output tokens} & \textbf{Cost} & \textbf{Time} \\ \midrule
\multirow{3}{*}{Tuning} & Stack & 5 & 189K & 9K & \$0.46 & 80s \\
 & Queue & 5 & 605K & 12K & \$0.72 & 143s \\
 & Hash table & 5 & 475K & 10K & \$0.60 & 106s \\
\midrule
\multirow{2}{*}{Held-out} & 2-lock queue & 4 & 1232K & 16K & \$1.04 & 176s \\
 & Harris list & 4 & 672K & 13K & \$0.81 & 222s \\
\bottomrule
\end{tabular}
\captionlabel{Evaluation of \logic being used by Claude Opus 5.5 to prove non-linearizability. %
The numbers are averaged over the cases of each data structure and over 3 runs. %
Between runs, the cost of a case varies by $\pm8\%$ on average (at most $\pm19\%$). %
Per case, the cost ranges from \$0.43 to \$1.10 (average \$0.71).}{fig:nonlin_res:summary}
\end{figure}

\Cref{fig:nonlin_res:summary} presents the results of our evaluation.
Each line groups the cases of one data structure, along with statistics about token costs and time;
\Appref{sec:nonlin_res} details each case and the nature of its bug.
The cost is \$0.71 per case on average, and at most \$1.10.
On average, the held-out data structures are the most expensive
(\$1.04 per case for the two-lock queue and \$0.81 for Harris's list).
In particular, the two-lock queue relies on locks,
which the proof guide does not mention,
as for the held-out memory race cases~(\cref{sec:study:race}).
Across all cases, about 30\% of the wall-clock time is spent
waiting for Rocq, rising to 50\% for Harris's list because
of the large goal sizes it generates.
Concerning Rocq tooling, the agent needs few attempts:
about 10 Rocq calls per case, of which 1.6 fail on average,
typically a tactic applied too early.
The costlier detours come from coverage gaps.
The two-lock queue projects fields out of an immutable tuple,
for which no tactic is documented.
Each run found its own workaround by applying
a generic lemma located with \texttt{Search}.
Better tactic coverage and faster proof checking
therefore seem to be the main levers to reduce cost.

\subsection{Finding Linearizability Bugs in Saturn}
\label{sec:study:saturn}

The experiments described so far have focused on running Claude Opus 5.5
equipped with
Mizzle over examples that were designed to be buggy.
We also launched an agent on the Saturn library~\citep{saturn}
version \texttt{1.0.0}, a general-purpose lock-free data structure library for OCaml.
The agent was given the non-linearizability proof guide~(\cref{sec:study:methodo}),
all the already-verified examples,
and tasked with finding linearizability bugs in the library,
as well as to use Mizzle to certify them.
The agent found two bugs, developed a small script reproducing the bugs,
and certified both bugs with Mizzle.
We subsequently reported the two bugs to the authors of the library.

First, the agent found
a non-linearizable execution of an \texttt{is\_full} query in a
bounded variant of the Michael and Scott queue~\citep{michael-scott-96}.
While an insertion has linked its node but not yet moved the tail,
\texttt{is\_full} trusts the lagging tail and can return \texttt{false} on a full queue.

Second, the agent found a non-linearizable execution of the \texttt{length} function
of both the
hash table and the
skiplist,
which share a size counter adapted from \citet{sela-petrank-22}.
An operation that helps a pending insertion
may skip forwarding it to a concurrently executing \texttt{length},
which then observes a later removal but not the insertion,
and returns a wrong size.
This bug is particularly interesting as it is hard to reproduce,
because the triggering window is small.
With a dedicated script where
one thread loops inserting the same key,
a second thread loops deleting this key,
and a third one loops computing the size,
the bug shows in around 1 in 20,000 runs.

\section{Related Work}
\label{sec:related_work}
\paragraph{Under-approximate logics}
The idea of an under-approximate logic,
that is, a logic whose triples describe a
\emph{subset} of the reachable behaviors
rather than an over-approximating superset,
is generally credited to \citet{devries-koutavas-11}, who introduce
\emph{reverse Hoare logic} to reason about reachability in nondeterministic
programs.
\citet{ohearn-il-20} develops \emph{Incorrectness Logic} (IL),
observing that under-approximation is exactly what is needed to justify a
bug-finding analysis that reports only true positives: an IL triple
$[\mathit{pre}]\,c\,[\mathit{post}]$ guarantees that every state in
$\mathit{post}$ is reachable from some state in $\mathit{pre}$.
\citet{raad-isl-20} first bring the power of
separation logic to IL and proposes
\emph{Incorrectness Separation Logic} (ISL), recovering the local, frame-based
reasoning that makes separation logic scalable while soundly detecting
memory-safety bugs.

\citet{raad-et-al-22}
extend under-approximate reasoning to
concurrency with
\emph{Concurrent Incorrectness Separation Logic} (CISL),
a parametric framework designed to underpin
compositional bug-finding tools such as RacerD~\citep{blackshear-et-al-18}.
CISL distinguishes three categories of concurrency bugs,
and detects only the first two:
local bugs, which manifest without concurrency;
global data-agnostic bugs, which manifest only under some interleavings,
but not because threads exchange information;
and global data-dependent bugs, which arise from such an exchange.
\citet{raad-et-al-23} generalize CISL with
\emph{Concurrent Adversarial Separation Logic} (CASL),
designed to detect software exploits
by encoding an adversarial client as a thread running in parallel with the vulnerable program.
CASL also addresses global data-dependent bugs,
thanks to rules inspired by rely-guarantee reasoning~\citep{jones-83}
and by recording the interleavings each thread assumes.
Both logics aim at \emph{thread-modular} reasoning:
bugs are detected by reasoning about each thread in isolation,
abstracting over interleavings.

\logic addresses a different problem:
certifying a bug proposed by an external party.
What matters then is that the logic can follow any interleaving,
compositionally.
\logic achieves this with rules that follow the operational semantics
while mentioning only the thread it steps and the memory it accesses,
the rest of the context being framed~(\cref{sec:key:logic}).
As a result, \logic handles all three categories without rely-guarantee reasoning,
allows the user to reason about threads independently,
and is proved complete.
Neither CISL nor CASL claims completeness.
To our knowledge, \logic is the first under-approximate logic for
concurrency that is proved complete.
Its language has \texttt{fork}, functions, recursion, and higher-order functions,
whereas CISL/CASL have structured parallel composition and a fixed loop construct.
Our approach lets locks be ordinary code~(\appref{sec:appendix:lock}),
and non-linearizability be a library-level specification.
Conversely, \logic has no notion of deallocation (because \ocaml{} has no deallocation),
which CISL/CASL handle, allowing them to consider use-after-free and double-free bugs.

Under-approximate logics exist outside of the world of incorrectness.
For example, Angelic~\citep{moine-westrick-tassarotti-26}
is a logic allowing for proving that \emph{one} interleaving
is safe and terminates---such a property is used in combination
with another property asserting that the verification of a single interleaving suffices to guarantee that they are all correct.
Angelic comes with a WP-like proof mode with a \textsf{run} modality targeting a single thread at a time,
encoded on top of a ``scheduler'' mode, which our $\goal$ assertion generalizes~(\cref{sec:logic:goal}).
Besides, Angelic targets parallel programs whereas \logic targets fork-based concurrency,
and an Angelic proof goes until the end of a terminating execution,
whereas a \logic proof stops as soon as it reaches an incorrect state.
Some logics try to combine
under-approximate reasoning with over-approximate reasoning.
In particular, Outcome Logic~\citep{zilberstein-25} offers a framework
where both demonic choice (for over-approximation)
and angelic choice (for under-approximation) can be expressed.
Exact separation logic~\citep{maksimovic-et-al-23} also combines both under- and over-approximate reasoning,
requiring the postcondition to describe exactly the states reachable from the precondition.

\paragraph{Automated static analyses for finding true positives}
Many automated tools are designed to avoid false positives,
and the incorrectness logics described above have been used to justify that these analyses have this property.
For example, the Pulse analysis~\citep{le-et-al-22} shipped
in Infer~\citep{facebook-infer} is grounded in ISL.
On the data race side, the no-false-positives property of
RacerD~\citep{blackshear-et-al-18} is proven using CISL.
Other under-approximate reasoning is also used
in a deadlock detector~\citep{brotherston-et-al-21}.
More compositional symbolic execution engines
such as Gillian~\citep{gillianpartone,gillian}
and Soteria~\citep{soteria}
also allow for under-approximate reasoning about programs
written in a variety of languages.
Recently, \citet{bayer2026teachingllmsprogramsemantics}
use Soteria to train LLMs on traces.

\paragraph{Non-linearizability and its detection}
A number of tools exist for automatically proving linearizability~\citep{vafeiadis2010-cave, zhu2015-poling, meyer2023-nekton}
 or checking for non-linearizability~\citep{DBLP:conf/pldi/BurckhardtDMT10, DBLP:journals/jpdc/WingG93, DBLP:journals/concurrency/Lowe17, DBLP:conf/forte/HornK15a, DBLP:journals/pacmpl/EmmiE18}.
Although checking whether a trace is non-linearizable is NP-hard, these tools develop approaches that scale well to realistic use cases.
Most non-linearizability checkers are designed to avoid false alarms, but their ability to detect bugs is bounded by the traces or test depths they explore.

\paragraph{LLMs writing machine-checked proofs}
There is a vast and growing literature on using LLMs to construct machine-checked proofs with proof assistants~\citep{alphaproof, goedelproverv2, deepseekproverv2} or to produce annotations for auto-active deductive program verifiers~\citep{alphaverus, autoverus, laurel, dafnypro}.
To the best of our knowledge, our work is the first to propose using LLMs to construct proofs of \emph{incorrectness} of programs.

\section{Discussion and Future Work}
\label{sec:conclusion}

We presented \logic,
a sound and complete
incorrectness separation logic
for a subset of \ocaml{} programs,
mechanized in Rocq using the
Iris framework. %
We now discuss avenues for future work.

\paragraph{Towards a full-fledged tool}
Our Saturn case study (\cref{sec:study:saturn}) shows that
\logic can certify bugs in unmodified source code;
we would like to scale this to a systematic study of \ocaml{} libraries,
starting from bug reports generated by LLMs.
We would also like to axiomatize functions that are expensive
or difficult to run during a test,
such as functions interacting over the network.
Testing usually replaces them by \emph{mock} functions,
which may not exhibit the full space of behaviors of the real environment,
and hence miss bugs;
a \logic axiom can instead capture the full set of allowable behaviors.
Finally, \logic assumes sequential consistency;
we would like to support the \ocaml{} memory model,
for example following the approach of Cosmo~\citep{cosmo},
in order to certify bugs whose symptoms only appear under weak memory.

\paragraph{Reusing a correctness proof}
A \emph{correctness} specification can be reused in \logic{}:
if we have a weakest precondition specification for a program $\expr$ of the form
$\pre \wand \wp\expr{\post}$,
where $\pre$ is the precondition and $\post$ is the postcondition,
then this guarantees that,
starting from a state satisfying~$\pre$,
\emph{every} interleaving of $\expr$ is safe (\ie, has no bug),
and if it terminates, it reaches a state satisfying $\post$.
Hence, if $\expr$ terminates,
this guarantees that \emph{at least one}
interleaving of $\expr$ can reach a state satisfying~$\post$ from a state satisfying $\pre$.
Thus, as preliminary mechanized experiments suggest,
if we use a \emph{total} weakest precondition,
which implies termination, it should also imply a \logic specification.
This would give us a way to intertwine the use of \logic with formal verification:
if a program is verified, then we can reuse its specification in \logic to find bugs in
clients being developed.
We would like to explore this further.

\begin{acks}
This work was supported in part by the \grantsponsor{NSF}{National Science Foundation}{} under Grant No.~\grantnum{NSF}{2524669}.
Any opinions, findings, and conclusions or recommendations expressed in this material are those
of the authors and do not necessarily reflect the views of this funding agency.
\end{acks}

\section*{Data-Availability Statement}
An artifact is available~\citep{anonymous-supplementary-material}.
This artifact includes an appendix to the present paper,
as well as the implementation of Mizzle in Rocq,
covering the logic itself, its soundness and completeness proofs, and our case studies.
Moreover, the artifact contains the logs of the agents used in our case studies.
Upon acceptance, we will make the artifact publicly available and submit it for artifact evaluation.

Additionally, in \Cref{sec:study:saturn},
we refer to bug reports made to the authors of Saturn~\citep{saturn}.
These reports are publicly available on GitHub and we will link them
in the paper once the double-blind period is over.
\bibliography{main}

\ifshort\else

\appendix

\section{Semantics of \lang}
\label{sec:appendix:semantics}

In this section, we complete the presentation of \lang given in \cref{sec:semantics}.
We first give some additional head reduction rules,
for blocks and memory accesses~(\cref{fig:semantics:mem}).
We then explain a subtle difference with \ocaml,
the fact that our semantics makes \oob accesses stuck~(\cref{sec:appendix:oob}).
Finally, we list the (presentational) differences between the language presented in this paper
and the one of our mechanization~(\cref{sec:appendix:differences}).

\subsection{Rules for Blocks and Memory Accesses}
\begin{figure}[t]\centering\small\morespacingaroundstar

\begin{mathpar}
\inferrule[Block]
  {0 < \lvert\vec\val\rvert \\ \disjoint{\big(\{\loc\} \cup \{\loc + \ofs \mid 0 \leq \ofs < \lvert\vec\val\rvert\}\big)}{\big(\dom(\tagmap) \cup \dom(\store)\big)}}
  {\headstep{\eblock\tagn{\vec\val}}{\tagmap}{\store}{\trace}{\loc}{\mapupd{\loc}{\headerval{\tagn}{\lvert\vec\val\rvert}}{\tagmap}}{\mapupd{\loc + \ofs}{\vec\val(\ofs) \mid 0 \leq \ofs < \lvert\vec\val\rvert}{\store}}{\trace}{\epsilon}}

\inferrule*[Left=CASSucc]
  {\store(\loc + \ofs) = \val \\ \val \approx \val_1 \\ \tagmap(\loc) = \headerval{\tagn}{m} \\ 0 \leq \ofs < m}
  {\headstep{\ecas{(\loc, \ofs)}{\val_1}{\val_2}}{\tagmap}{\store}{\trace}{\vtrue}{\tagmap}{\mapupd{\loc + \ofs}{\val_2}{\store}}{\trace}{\epsilon}}

\inferrule*[Left=CASFail]
  {\store(\loc + \ofs) = \val \\ \val \napprox \val_1 \\ \tagmap(\loc) = \headerval{\tagn}{m} \\ 0 \leq \ofs < m}
  {\headstep{\ecas{(\loc, \ofs)}{\val_1}{\val_2}}{\tagmap}{\store}{\trace}{\vfalse}{\tagmap}{\store}{\trace}{\epsilon}}
\end{mathpar}
\captionlabel{Further rules of the semantics for block construction and compare-and-set}{fig:semantics:mem}
\end{figure}

\Cref{fig:semantics:mem} presents further head reduction rules of \lang
for block construction and compare-and-set,
complementing \cref{fig:semantics}.
\RULE{Block} allocates a fresh block;
unlike \RULE{Alloc}, its size is the number of its fields, which must be non-zero.
\RULE{CASSucc} and \RULE{CASFail} perform a compare-and-set
on the field at offset~$\ofs$ of the block at location~$\loc$.
Like \RULE{Load} and \RULE{Store}, they check that $\ofs$ is within bounds
using the header of the block.
They compare the current value~$\val$ of the field with the expected value~$\val_1$
using physical equality~(\cref{sec:key:zoo}):
if $\val \approx \val_1$, \RULE{CASSucc} stores $\val_2$ and returns $\vtrue$;
if $\val \napprox \val_1$, \RULE{CASFail} leaves the heap unchanged and returns $\vfalse$.
As $\approx$ and $\napprox$ are not inverse to each other,
both rules may apply to the same configuration.

\subsection{Out-of-Bound Accesses}
\label{sec:appendix:oob}
\Oob accesses are a source of bugs
that are not prevented by OCaml's type system.
As a mitigation, OCaml's standard library functions
usually come in two flavors:
a default version testing that the accessed
offset is within bounds, raising an exception otherwise,
and an optimized version bypassing the check.
For example, the \ocamli{Array} module
of the OCaml standard library
offers two primitives to load the content of
an offset within an array:
\ocamli{Array.get} and \ocamli{Array.unsafe_get}.

\logic is able to catch \oob errors,
whether they are checked at runtime
(in this case, looking for a stuckness bug suffices),
or not (since such errors can be encoded as predicates over configurations, see \Cref{sec:defining_incorrectness}).
However, while an \oob access
can lead to a segmentation fault
and the termination of the execution,
it can sometimes land on another allocated object,
read some data, and silently proceed.
\logic is not able to catch bugs that are reached after such
a successful \oob read.
Indeed, the \RULE{M-Block}
rule gives no control over the universally-quantified location,
hence it is not possible to force the alignment of memory blocks.
Arguably, a bug that can only be reached after a successful
\oob access is not the priority---the \oob access is.
Hence, the reduction relation $\step{\config_1}{\config_2}$ of \cref{fig:semantics,fig:semantics:mem}
prevents \oob accesses: the field-access rules guard every access with an
in-bounds side condition (\eg, $0 \leq \ofs < m$ in \RULE{Load}).

\subsection{Differences with the Mechanization}
\label{sec:appendix:differences}
For exposition purposes,
this paper presents a subset of \lang~(\cref{sec:syntax}); the
mechanized development uses the full language, which differs as follows.
First, we omit \emph{immutable blocks}: in the mechanization, a block
carries a mutability flag, and an immutable block is not allocated on the heap
but evaluates, after a single step, to a value carrying its tag and fields.
Loads from immutable blocks need no points-to assertion,
and physical equality on them is underspecified~(\cref{sec:key:zoo}).
We also omit \emph{generativity}, a mechanism that gives immutable
blocks an observable identity.
Second, \lang functions are in fact \emph{mutually recursive} bundles
together with a selector index; we ignore this aspect.
Third, we omit the $\textsf{match}$ construct, which inspects a block
by its tag and binds its fields.
Fourth, we omit the atomic operations $\textsf{Xchg}$ (exchange) and
$\textsf{FAA}$ (fetch-and-add).
Fifth, we omit the bounded $\textsf{for}$ loop and the $\textsf{while}$ loop,
which are otherwise standard.
Sixth, we omit thread-local state, accessed in \lang through dedicated
$\textsf{getLocal}$ and $\textsf{setLocal}$ primitives.
Seventh, the mechanization does not store the trace in the configuration:
\RULE{Emit} leaves the state unchanged and instead labels its reduction step
with the emitted value, and the trace is the concatenation of the labels
along a reduction path.
The two presentations are equivalent.

\section{Detailed Case Studies Results}
In this section, we complete the presentation
of the case studies of \cref{sec:case_studies}.
We first present the spin lock used by some cases~(\cref{sec:appendix:lock}).
We then give the detailed results of each run for memory races~(\cref{sec:drb_res})
and comment on representative examples of programs
we prove to have a memory race~(\cref{sec:appendix:races}).
We finally give the detailed results of each run
for non-linearizability~(\cref{sec:nonlin_res}).

\subsection{A Spin Lock}
\label{sec:appendix:lock}
\begin{figure}
\centering
{\small
\[\begin{array}{@{}l@{\;\eqdef\;}l@{}}
\textsf{create} & \vrec{\_}{\_}{\eblock{0}{[\vfalse]}} \\
\textsf{lock} & \vrec{\textsf{lock}}{t}{\eif{\eunop\sim{(\ecas{\vtuple{t, 0}}\vfalse\vtrue)}}{\textsf{lock}\,t}{\vunit}} \\
\textsf{unlock} & \vrec{\_}{t}{\estore[\matomic]{t}{0}{\vfalse}}
\end{array}\]}

\bigskip
{\small\morespacingaroundstar\morespacingaroundvdash
\[\begin{array}{@{}c@{\quad}r@{\;\vdash\;}l@{}}
\TirNameStyle{M-LockCreate} & \mlprec{
\thread\tid\ectxs{(\textsf{create}\,\vunit)} \star {}\\
(\forall \loc.\; \cells\loc{[\vfalse]} \star \thread\tid\ectxs\loc \wand \goal)
}
  & \goal \\[2.5ex]

\TirNameStyle{M-Lock} & \mlprec{
\thread\tid\ectxs{(\textsf{lock}\,\loc)} \star \cells\loc{[\vfalse]} \star {}\\
(\cells\loc{[\vtrue]} \star \thread\tid\ectxs\vunit \wand \goal)
}
  & \goal \\[2.5ex]

\TirNameStyle{M-Unlock} & \mlprec{
\thread\tid\ectxs{(\textsf{unlock}\,\loc)} \star \cells\loc{[\val]} \star {}\\
(\cells\loc{[\vfalse]} \star \thread\tid\ectxs\vunit \wand \goal)
}
  & \goal
\end{array}\]}
\captionlabel{A spin lock and its specification in \logic}{fig:spinlock}
\end{figure}

The held-out memory race cases with missing synchronization
use locks, which we implement as spin locks.

The code appears in the upper part of \Cref{fig:spinlock}.
A lock is a mutable block (with tag 0)
with a single boolean field, $\vfalse$ when unlocked
and $\vtrue$ when locked.
The function $\textsf{lock}$ loops until its compare-and-set succeeds,
and $\textsf{unlock}$ releases the lock with an atomic store.
Since every access to the cell is atomic,
the lock itself is never part of a memory race.

The lower part of \Cref{fig:spinlock}
presents the \logic specifications we provide and verify for this lock.
\RULE{M-LockCreate} hands the user a fresh location~$\loc$
with a points-to $\cells\loc{[\vfalse]}$ for an unlocked lock.
\RULE{M-Lock} consumes the points-to of an unlocked lock
and produces the points-to of a locked lock.
The fact that \RULE{M-Lock} applies only to an unlocked lock is enough for bug finding:
since \logic lets the user choose the interleaving,
a thread trying to acquire a held lock is simply not scheduled
until the lock is released.
\RULE{M-Unlock} requires the points-to of the lock, whatever its content~$\val$,
and gives back the points-to of an unlocked lock.

\subsection{Memory Races}
\label{sec:drb_res}
\begin{figure}
\centering\small
\begin{tabular}{llllrrrr}
\toprule
\textbf{Set} & \textbf{ID} & \textbf{Family} & \textbf{Bug} & \makecell[r]{\textbf{Input}\\\textbf{tokens}} & \makecell[r]{\textbf{Output}\\\textbf{tokens}} & \textbf{Cost} & \textbf{Time} \\ \midrule
\multirow{12}{*}{\makecell[l]{Tuning}} & 009 & Shared cell & Shared write & 176K & 3K & \$0.25 & 27s \\
 & 011 & Shared cell & Shared decrement & 581K & 11K & \$0.64 & 109s \\
 & 016 & Shared cell & Write-after-write & 165K & 3K & \$0.27 & 35s \\
 & 018 & Shared cell & Shared increment & 194K & 4K & \$0.29 & 39s \\
 & 020 & Shared cell & Shared temporary & 197K & 5K & \$0.32 & 42s \\
 & 003 & Loop-carried & Write-after-read (2D loop) & 402K & 6K & \$0.57 & 64s \\
 & 005 & Loop-carried & Indirect indexing & 730K & 14K & \$0.81 & 144s \\
 & 006 & Loop-carried & Indirect indexing & 360K & 10K & \$0.54 & 99s \\
 & 014 & Loop-carried & Cross-row read (j-1 at j=0) & 264K & 5K & \$0.36 & 51s \\
 & 015 & Loop-carried & Cross-row read (var. length) & 223K & 5K & \$0.34 & 48s \\
 & 024 & Loop-carried & Read-after-write (two arrays) & 201K & 6K & \$0.38 & 55s \\
 & 029 & Loop-carried & Read-after-write & 154K & 4K & \$0.29 & 37s \\
\midrule
\multirow{22}{*}{\makecell[l]{Held-\\out}} & 039 & Shared cell & Read of a[0] while iter. 0 writes & 165K & 3K & \$0.27 & 33s \\
 & 111 & Shared cell & Shared index counter & 309K & 8K & \$0.46 & 77s \\
 & 169 & Shared cell & Shared scratch array & 323K & 7K & \$0.46 & 88s \\
 & 202 & Shared cell & Read of a[64] while iter. 64 writes & 169K & 4K & \$0.31 & 46s \\
 & 033 & Loop-carried & Read-after-write (affine index) & 168K & 4K & \$0.31 & 39s \\
 & 037 & Loop-carried & Read-after-write (in seq. loop) & 492K & 7K & \$0.51 & 70s \\
 & 179 & Loop-carried & Write-after-write (conditional) & 284K & 4K & \$0.35 & 44s \\
 & 195 & Loop-carried & Stencil on aliased buffers & 870K & 17K & \$1.00 & 210s \\
 & 197 & Loop-carried & Overlapping buffers (fixed offset) & 598K & 13K & \$0.71 & 127s \\
 & 204 & Loop-carried & Affine read (2i vs. i+64) & 289K & 7K & \$0.43 & 71s \\
 & 206 & Loop-carried & Affine write (2i vs. i+64) & 304K & 7K & \$0.45 & 67s \\
 & 207 & Loop-carried & Read-after-write (distance 64) & 316K & 8K & \$0.45 & 73s \\
 & 074 & Missing sync. & Read outside the lock & 276K & 4K & \$0.33 & 42s \\
 & 119 & Missing sync. & Increment outside the lock & 232K & 3K & \$0.29 & 35s \\
 & 124 & Missing sync. & Missing barrier after write & 162K & 2K & \$0.23 & 24s \\
 & 160 & Missing sync. & Missing barrier between loops & 345K & 9K & \$0.51 & 85s \\
 & 185 & Missing sync. & Broken custom barrier (flags) & 548K & 11K & \$0.68 & 107s \\
 & 187 & Missing sync. & Broken lock-based barrier (2 locks) & 262K & 4K & \$0.32 & 42s \\
 & 189 & Missing sync. & Broken lock-based barrier (3 locks) & 543K & 13K & \$0.70 & 157s \\
 & 191 & Missing sync. & Distinct critical sections & 235K & 4K & \$0.31 & 42s \\
 & 193 & Missing sync. & Distinct critical sections (flag) & 234K & 4K & \$0.31 & 42s \\
 & 201 & Missing sync. & Write after empty critical section & 232K & 3K & \$0.29 & 34s \\
\bottomrule
\end{tabular}
\captionlabel{Evaluation of \logic being used by Claude Opus 5.5 to prove the existence of a memory race. %
The numbers are averaged over 3 total runs. %
Between runs, the cost of a case varies by $\pm5\%$ on average (at most $\pm21\%$).}{fig:drb_res}
\end{figure}

\Cref{fig:drb_res} presents
the detailed results of our evaluation of \logic
on the memory race cases.

\subsection{Examples of Memory Races}
\label{sec:appendix:races}
\begin{figure}
\newcommand{\drbfontsize}{\small}
\centering
\begin{minipage}[t]{.42\textwidth}
\begin{subfigure}[t]{\textwidth}
\centering
\begin{minted}[fontsize=\drbfontsize]{ocaml}
let main () =
  let a = Array.make 100 0 in
  let x = ref 10 in
  parallel_for 100
    (fun i -> a.(i) <- !x; x := i)
\end{minted}
\medskip
\captionlabel{\textbf{Shared cell} (016): concurrent reads and writes to the
shared cell \texttt{x}}{fig:drb_examples:shared}
\end{subfigure}

\bigskip\bigskip
\begin{subfigure}[t]{\textwidth}
\centering
\begin{minted}[fontsize=\drbfontsize]{ocaml}
let main () =
  let n = 100 in
  let a = Array.make n 0 in
  for i = 0 to n - 1 do
    a.(i) <- i
  done;
  parallel_for (n - 1)
    (fun i -> a.(i+1) <- a.(i) + 1)
\end{minted}
\medskip
\captionlabel{\textbf{Loop-carried} (029): iteration \texttt{i} reads
\texttt{a.(i)} written by iteration \texttt{i-1}}{fig:drb_examples:carried}
\end{subfigure}
\end{minipage}%
\hfill
\begin{minipage}[t]{.55\textwidth}
\begin{subfigure}[t]{\textwidth}
\centering
\begin{minted}[fontsize=\drbfontsize]{ocaml}
let n = 180
let main () =
  let index_set = Array.make n 0 in
  for g = 0 to 29 do
    let start = 521 + 338*g/6 + 26*(g mod 6) in
    for k = 0 to 5 do
      index_set.(6*g + k) <- start + 2*k
    done
  done;
  index_set.(5) <- 533;
  let base = Array.make (2013+12+1) 0 in
  for i = 521 to 2025 do base.(i) <- i done;
  parallel_for n (fun i ->
    let idx = index_set.(i) in
    base.(idx)    <- base.(idx)    + 1;
    base.(idx+12) <- base.(idx+12) + 3)
\end{minted}
\medskip
\captionlabel{\textbf{Loop-carried (ind.)} (006): the offset \texttt{idx} is computed
at runtime, opening the door to collisions}{fig:drb_examples:indirect}
\end{subfigure}
\end{minipage}
\vspace{0.75em}
\captionlabel{Three representative examples: a shared cell, a
loop-carried dependence, and a loop-carried dependence through runtime-computed offsets}{fig:drb_examples}
\end{figure}

\paragraph{Representative examples}
\Cref{fig:drb_examples} presents three racy examples
from our tuning set.
We comment on the logs of the three runs of each of them.
In all nine runs of these three examples, the agent starts a Rocq session,
writes the whole proof by adapting the canonical proof of the proof guide,
and finally checks the proof file.

\Cref{fig:drb_examples:shared} shows a shared-cell bug:
while the ownership of the array \ocamli{a} is correctly
split for the parallel for loop,
all iterations are accessing the same shared cell \ocamli{x} for both reading and writing.
In each run, the agent identifies that the race is on \ocamli{x} and not on \ocamli{a},
and its first attempt at the whole proof succeeds,
in 30 to 42 seconds and for less than \$0.28.
It parks one thread at its read of \ocamli{x},
and the other at its write to \ocamli{x}.
The main technical difficulty the agents reported was
the right-to-left evaluation order,
because of which \ocamli{!x} is evaluated before the write to \ocamli{a}.

\Cref{fig:drb_examples:carried} shows a loop-carried bug:
iterations of the parallel loop do have separate ownership,
as iteration \ocamli{i} writes to the \ocamli{(i+1)}-th
offset of the array \ocamli{a}, while reading offset \ocamli{i},
which is being concurrently written by iteration \ocamli{i-1}.
Again, the first attempt at the whole proof succeeds in each run,
in about 37 seconds and for less than \$0.30.
The agent reports that this case closely follows the canonical proof,
with which it shares the initialization loop:
it only changes the sizes,
and swaps which thread is parked at the write and which at the read.

\Cref{fig:drb_examples:indirect} shows a more
complex case: the program first
allocates an array of offsets \ocamli{index_set},
populates it with complex calculations, but with an injective function,
next overwrites one of the offsets to create a collision,
and then allocates an array \ocamli{base},
before running a parallel for loop,
accessing offsets in \ocamli{base} depending on the values in \ocamli{index_set}.
Each run takes 92 to 106 seconds and costs less than \$0.56,
with five calls to the Rocq checker.
The agent works out the arithmetic
to understand which offset has a race:
iteration~0 (where \ocamli{idx=521}) and iteration 5 (where \ocamli{idx=533})
both touch the same offset \ocamli{533=521+12}.
The main challenge the agent reports is that
a loop invariant that only tracks the ownership of the cells
forgets the value of \ocamli{index_set.(0)}.
In all three runs, it adopts the same solution:
execute first the iteration of the nested initialization loops by hand,
which records that \ocamli{index_set.(0)} contains \ocamli{521},
and then executes the remaining iterations
with an invariant that only owns the cells $1$ to $179$,
since no later iteration writes to cell $0$.
In two of the three runs, one call to the Rocq checker fails:
once because the agent tries to step a read before the statement preceding it,
and once because it passes evaluation contexts
instead of thread identifiers to a lemma.

\paragraph{Case 011}
Let us finally comment on case 011,
which reaches the highest rate of MCP failures in the tuning set.
The code is as follows:
\begin{minted}[fontsize=\small]{ocaml}
let main () =
  let len = 100 in
  let num_nodes = len in
  let num_nodes2 = ref 0 in
  let x = Array.make len 0 in
  for i = 0 to len - 1 do if i mod 2 = 0 then x.(i) <- 5 else x.(i) <- -5 done;
  parallel_for num_nodes (fun i -> if x.(i) <= 0 then num_nodes2 := !num_nodes2 - 1)
\end{minted}
The race occurs because multiple offsets of \ocamli{x} store negative values,
passing the test \ocamli{x.(i) <= 0} and hence concurrently decrementing \ocamli{num_nodes2}.
In each of the three runs, the agent immediately identifies the race,
between iterations 1 and 3.
To prove that both iterations pass the test,
it must know that \ocamli{x.(1)} and \ocamli{x.(3)} are negative
after the initialization loop:
again, an invariant that only owns the cells is not enough.
Each run adopts a different strategy, executing different portions
of the loop by hand.
Moreover, the test \ocamli{i mod 2 = 0} uses
OCaml's \emph{structural} equality,
which recurses through immutable data structures,
as opposed to physical equality~(\Cref{sec:key:zoo}).
In \lang, structural equality is implemented as a library function,
not as a primitive.
The three runs respectively make 2, 5 and 10 failed calls to the Rocq checker.
Besides structural equality, failures arise
because of Ltac, hallucination of a lemma,
trying to unfold an opaque definition,
consuming a hypothesis that was later needed,
and other technicalities.

\subsection{Non-Linearizability}
\label{sec:nonlin_res}
\begin{figure}
\centering\small
\begin{tabular}{llllrrrr}
\toprule
\textbf{Set} & \textbf{ID} & \textbf{Structure} & \textbf{Bug} & \makecell[r]{\textbf{Input}\\\textbf{tokens}} & \makecell[r]{\textbf{Output}\\\textbf{tokens}} & \textbf{Cost} & \textbf{Time} \\ \midrule
\multirow{15}{*}{\makecell[l]{Tuning}} & 001 & Stack & Lost push (non-atomic store) & 177K & 9K & \$0.46 & 83s \\
 & 002 & Stack & Double pop (non-atomic store) & 197K & 8K & \$0.45 & 71s \\
 & 003 & Stack & Clobbering push (fresh-read CAS) & 161K & 8K & \$0.43 & 77s \\
 & 004 & Stack & Lost push (discarded CAS) & 195K & 8K & \$0.44 & 78s \\
 & 005 & Stack & Spurious empty pop (no retry) & 216K & 10K & \$0.51 & 92s \\
 & 006 & Queue & Lost dequeue (no retry) & 916K & 16K & \$0.95 & 177s \\
 & 007 & Queue & Double dequeue (blind store) & 316K & 9K & \$0.52 & 95s \\
 & 008 & Queue & Double dequeue (fresh-read CAS) & 441K & 10K & \$0.59 & 114s \\
 & 009 & Queue & Lost enqueue (discarded CAS) & 932K & 15K & \$0.87 & 167s \\
 & 010 & Queue & Double dequeue (stale peek) & 423K & 12K & \$0.67 & 164s \\
 & 011 & Hash table & Lost add (blind store, add) & 445K & 8K & \$0.52 & 84s \\
 & 012 & Hash table & Lost add (blind store, remove) & 611K & 14K & \$0.80 & 153s \\
 & 013 & Hash table & Lost add (fresh-read CAS) & 357K & 8K & \$0.54 & 95s \\
 & 014 & Hash table & Lost add (CAS-fail reports ok) & 498K & 9K & \$0.57 & 92s \\
 & 015 & Hash table & Double add (stale lookup) & 464K & 9K & \$0.58 & 105s \\
\midrule
\multirow{8}{*}{\makecell[l]{Held-\\out}} & 016 & 2-lock queue & Lost enqueue (no lock) & 1302K & 17K & \$1.10 & 180s \\
 & 017 & 2-lock queue & Double dequeue (unlocked read) & 1346K & 16K & \$1.08 & 180s \\
 & 018 & 2-lock queue & Spurious empty dequeue (late link) & 1019K & 15K & \$0.92 & 156s \\
 & 019 & 2-lock queue & Double dequeue (split section) & 1261K & 17K & \$1.05 & 188s \\
 & 020 & Harris list & Lost add (no logical deletion) & 673K & 14K & \$0.83 & 261s \\
 & 021 & Harris list & Removed key visible (mark ignored) & 857K & 16K & \$1.01 & 299s \\
 & 022 & Harris list & Double remove (unchecked mark) & 634K & 12K & \$0.74 & 211s \\
 & 023 & Harris list & Lost add (retry on deleted node) & 523K & 9K & \$0.65 & 116s \\
\bottomrule
\end{tabular}
\captionlabel{Evaluation of \logic being used by Claude Opus 5.5 to prove non-linearizability. %
The numbers are averaged over 3 total runs. %
Between runs, the cost of a case varies by $\pm8\%$ on average (at most $\pm19\%$).}{fig:nonlin_res}
\end{figure}

\Cref{fig:nonlin_res} presents
the detailed results of our evaluation of \logic
on the non-linearizability cases.
\fi

\end{document}